\documentclass{aa}
\usepackage{graphicx}
\usepackage{txfonts}
\usepackage{natbib}
\usepackage{color}
\bibpunct{(}{)}{;}{a}{}{,}    
\newcommand{\beq}{\begin{equation}}
\newcommand{\eeq}{\end{equation}}
\newcommand{\bea}{\begin{eqnarray}}
\newcommand{\eea}{\end{eqnarray}}

\newcommand{\pdoverdt}[1]{\frac{\partial #1}{\partial t}}
\newcommand{\subscr}[1]{_\mathrm{#1}}

\newcommand{\lsim}{\raisebox{-0.6ex}{$\stackrel{{\displaystyle<}}{\sim}$}}
\newcommand{\gsim}{\raisebox{-0.6ex}{$\stackrel{{\displaystyle>}}{\sim}$}}

\newcommand{\url}[1]{{\tt #1}}

\newcommand{\doverd}[2]{\frac{\partial #1}{\partial #2}}
\newcommand{\doverdt}[1]{\frac{\partial #1}{\partial t}}
\def\gapp{\lower 3pt\hbox{${\buildrel > \over \sim}$}\ }
\def\lapp{\lower 3pt\hbox{${\buildrel < \over \sim}$}\ }

\newlength{\linwx}
\begin{document}
\title{Planet migration in three-dimensional radiative discs}
\author{
Wilhelm Kley  \inst{1},
Bertram Bitsch \inst{1}
\and
Hubert Klahr \inst{2}
}
\offprints{W. Kley,\\ \email{wilhelm.kley@uni-tuebingen.de}}
\institute{
     Institut f\"ur Astronomie \& Astrophysik, 
     Universit\"at T\"ubingen,
     Auf der Morgenstelle 10, D-72076 T\"ubingen, Germany
\and
Max-Planck-Institut 
f\"ur Astronomie, Heidelberg
}
\date{February 25, 2009, revised May 27, 2009}
\abstract
{The migration of growing protoplanets depends on the thermodynamics of the
ambient disc. Standard modelling, using locally isothermal discs, indicate in the low
planet mass regime an inward (type-I) migration. Taking into account non-isothermal
effects, recent studies have shown that the direction of the type-I migration
can change from inward to outward.}
{In this paper we extend previous two-dimensional studies, and investigate the
planet-disc interaction in viscous, radiative discs using fully three-dimensional radiation
hydrodynamical simulations of protoplanetary accretion discs with embedded planets,
for a range of planetary masses.
}
{We use an explicit three-dimensional (3D) hydrodynamical code
{{\tt NIRVANA}} that includes full tensor viscosity.
We have added implicit radiation transport in the flux-limited diffusion
approximation, and to speed up the simulations significantly we have newly adapted
and implemented the {\tt FARGO}-algorithm in a 3D context.
}
{First, we present results of test simulations that demonstrate the accuracy of the
newly implemented {\tt FARGO}-method in 3D. For a planet mass of 20 $M_{\rm earth}$ we then
show that the inclusion of radiative effects yields a torque reversal also in full 3D.
For the same opacity law used the effect is even stronger in 3D than in the corresponding 2D
simulations, due to a slightly thinner disc. Finally, we demonstrate the extent of the
torque reversal by calculating a sequence of planet masses.
}
{Through full 3D simulations of embedded planets in viscous, radiative discs we confirm
that the migration can be directed outwards up to planet masses of about 33 $M_{\rm earth}$.
Hence, the effect may help to resolve the problem of too rapid inward migration of
planets during their type-I phase. 
}
\maketitle
\markboth
{Kley, Bitsch \& Klahr: Planet migration in radiative discs}
{Kley, Bitsch \& Klahr: Planet migration in radiative discs}

\section{Introduction}
\label{sec:introduction}
The process of migration in protoplanetary discs allows forming planets
to move away from their location of creation and finally end up at a
different position. The cause of this change in distance from the
star are the tidal torques acting from the disturbed disc back on the protoplanet.
These can be separated into two parts: {\it i)}
the so-called Lindblad torques that are created
by the two spiral arms in the disc and  {\it ii)} the corotation torques that are
caused by the co-orbital material as it periodically exchanges angular momentum
with the planet on its horseshoe orbit.
For comprehensive introductions to the field see for example
\citet{2007prpl.conf..655P} or \citet{2008EAS....29..165M}, and references therein. 
The Lindblad torques, caused by density waves launched at Lindblad resonances,
quite generally lead to an inward motion of the planet
explaining quite nicely the existence of the observed hot planets 
\citep{1997Icar..126..261W}.
The corotation torques are caused mainly by two effects, first by
a gradient in the vortensity \citep{2002ApJ...565.1257T}, and second by a
gradient in the entropy \citep{2008ApJ...672.1054B}.
For typical protoplanetary discs both contributions can
be positive, possibly counterbalancing the negative Lindblad torques 
\citep{2006A&A...459L..17P, 2008ApJ...672.1054B}.
For the typically considered locally isothermal discs where the temperature depends
only on the radial distance from the star the net torque is negative and migration
directed inwards for typical disc parameter \citep{2002ApJ...565.1257T}.
Through planetary synthesis models the inferred rapid inward migration
of planetary cores has been found
to be inconsistent with the observed mass-distance distribution of exoplanets
\citep{2004A&A...417L..25A, 2008ApJ...673..487I}. Possible remedies are the retention
of icy cores at the snow line or a strong reduction
in the speed of type-I migration (of embedded small mass planets).
Here we focus of the latter process.

Different mechanism for slowing down the too rapid inward migration
have been discussed \citep{2006ApJ...652..730M,2009ApJ...690L..52L},
but the inclusion of more realistic physics seems to be particularly appealing.
As has been pointed out first by \citet{2006A&A...459L..17P}
the inclusion of radiative transfer can cause a strong reduction in the migration
speed. 
The process has been subsequently investigated by several groups
\citep{2008ApJ...672.1054B, 2008A&A...478..245P,2008A&A...485..877P,2008A&A...487L...9K},
who show that indeed the migration process can be slowed down or even reversed for
sufficiently low mass planets. 
This new effect occurs in non-isothermal discs and scales with the gradient of the
entropy \citep{2008ApJ...672.1054B}, hence entropy-related torque. 
However, in a strictly adiabatic situation after a few libration time scales
the entropy gradient will flatten within the corotation region due to phase mixing.
This will lead to a saturation (and subsequent disappearance)
of the part of the corotation torque that is caused by the entropy gradient 
in the horseshoe region \citep{2008ApJ...672.1054B, 2009MNRAS.394.2283P}.
To prevent saturation of this entropy-related torque 
some radiative diffusion (or local radiative cooling) is required \citep{2008A&A...487L...9K}.
Since for genuinely inviscid flows, the streamlines in the horseshoe region will be closed
and symmetric with respect to the planet's location,
some level of viscosity is always necessary to avoid torque saturation
\citep{2001ApJ...558..453M, 2003ApJ...587..398O, 2008A&A...485..877P, 2009MNRAS.394.2283P}.
This applies to both, the vortensity- and entropy-related corotation torques.
The maximum planet mass for which a change of migration may occur due to this effect lies
for typical disc masses in the range of about
40 earth masses, beyond which the migration rate follows the standard (isothermal)
case, as gap formation sets in which reduces the corotation
effects \citep{2008A&A...487L...9K}. 
Most of the above simulations have studied only the two-dimensional case, while three
dimensional models including radiative effects have been presented only for
very small masses \citep{2006A&A...459L..17P}, or for Jupiter type planets
\citep{2006A&A...445..747K}. A range of planet masses  
has not yet been studied systematically in full 3D.

In this paper we investigate the planet-disc interaction in radiative discs
using fully three-dimensional radiation hydrodynamical simulations of protoplanetary
accretion discs with embedded planets for a variety of planetary masses.
For that purpose we modified and substantially extended an existing
multi-dimensional hydrodynamical code {\tt Nirvana}
\citep{1997ZiegYork,2001ApJ...547..457K} by incorporating the
{\tt FARGO}-algorithm \citep{2000A&AS..141..165M} 
and radiative transport in the flux-limited diffusion approximation \citep{1981ApJ...248..321L,1989A&A...208...98K}.
The code {\tt Nirvana} can in principle handle nested grids which allows to zoom-in on the detailed
structure in the vicinity of the planet \citep{2002A&A...385..647D,2003ApJ...586..540D},
however in the present context we limit ourselves to single grid simulations.
We present several test cases to demonstrate first the accuracy of the 
{\tt FARGO}-method in 3D. We then proceed to analyse the effects of radiative transport
on the disc structure and torque balance. For our standard planet of $20 M_{\rm earth}$,
we find that the effect of torque reversal appears to be
even stronger in 3D than in 2D for an otherwise identical physical setup.
We have a more detailed look at the implementation of the planet potential
and show that it has definitely an influence on the strength of the effect.
Finally, we perform simulations for a sequence of different planet masses to evaluate
the mass range over which the migration may be reversed. 
The consequence for the migration
process and the overall evolution of planets in discs is discussed.

\section{Physical modelling}
\label{sec:model}
The protoplanetary disc is treated as a three-dimensional (3D),
non-self-gravitating gas whose motion is described by the Navier-Stokes
equations. The turbulence in discs is thought to be driven
by magneto-hydrodynamical instabilities \citep{1998RvMP...70....1B}. Since we are
interested in this study primarily on the average effect the disc has on the planet,
we prefer in this work to simplify and treat the disc as a viscous medium.
The dissipative effects can then be described via the standard viscous stress-tensor approach
\citep[eg.][]{1984frh..book.....M}.
We assume that the heating of the disc occurs solely through
internal viscous dissipation and ignore in the present
study the influence of additional energy sources such as irradiation from the central star or
other external sources.
The internally produced energy is then radiatively diffused through the disc and eventually
emitted from its surfaces. To describe this process we utilise the flux-limited
diffusion approximation \citep[FLD,][]{1981ApJ...248..321L} which allows to treat approximately
the transition from optically thick to thin regions near the disc's surface.

\subsection{Basic equations}
Discs with embedded planets have mostly been modelled through 2D simulations
in which the disc is assumed to be infinitesimal thin,
and vertical integrated quantities are used to describe the time evolution
of the disc with the embedded planet. This procedure
saves considerable computational effort
but is naturally not as accurate as truly 3D simulations, in particular the
radiation transport is difficult to model in a 2D context.

In this work we present an efficient method for 3D disc simulations based on the
{\tt FARGO} algorithm \citep{2000A&AS..141..165M}. For accretion discs where material is
orbiting a central object the best suited coordinates are
spherical polar coordinates $(r, \theta, \varphi)$
where $r$ denotes the radial distance from the origin, $\theta$ the
polar angle measured from the $z$-axis, and $\varphi$ denotes the
azimuthal coordinate starting from the $x$-axis.

In this coordinate system, the mid-plane of the disc coincides with the
equator ($\theta=\pi/2$), and the origin of the coordinate system is
centred on the star.
Sometimes we will need the radial distance from the polar axis which
we denote by a lower case $s$, which is the radial coordinate in cylindrical
coordinates.

For a better resolution of the flow in the vicinity of the planet, we
work in a rotating coordinate system which rotates with the orbital
angular velocity $\Omega$, which is identical to the orbital angular
velocity of the planet
\begin{equation}
     \Omega_P = \left[\frac{G\,(M_*+m_p)}{a^3}\right]^{1/2}
\end{equation}
where $M_*$ is the mass of the star, $m_p$ the mass of the planet, and $a$
the semi-major axis of the planet.
Only for testing purposes for our implementation of
the {\tt FARGO}-method we let the planet move under the action of the disc.

The Navier-Stokes equations in a rotating coordinate system
in spherical coordinates read explicitly:

\noindent
a) Continuity Equation
\begin{equation}
 \doverd{\, \rho}{t}
   + \nabla \cdot (\rho\, {\bf u})
         = 0
      \label{eq:rho}
\end{equation}
Here $\rho$ denotes the density of the gas and 
${\bf u}=(u_r, u_\theta, u_\varphi)$ its velocity. 

\noindent
b) Radial Momentum
\begin{eqnarray}
 \doverdt{\, \rho u_r} + \nabla \cdot (\rho\,u_r\,{\bf u} )
  & = &  \rho \frac{u_\theta^2}{r} 
  +  \rho r \sin^2 \theta \, ( \omega + \Omega)^2  \nonumber \\
  & + &  \rho a_r
   -  \doverd{p}{r} - \rho\,\doverd{\Phi}{r} + f\subscr{r}. 
      \label{eq:u_r}
\end{eqnarray}
Here $\omega$ is the azimuthal 
angular velocity as measured in the rotating frame, 
$p$ is the gas pressure, and $\Phi$ denotes 
the gravitational potential due to the star and the planet.
The vector ${\bf a}=(a_r, a_\theta, a_\varphi)$ represents inertial
forces generated by the accelerated origin of the coordinate system.
Specifically, ${\bf a}$ equals the negative acceleration acting
on the star due to the planet(s). 

\noindent
c) Meridional Momentum
\begin{eqnarray}
 \doverd{\, \rho r u_\theta}{t}
   + \nabla \cdot (\rho r u_\theta {\bf u})
      &   = & 
     \rho r^2 \sin \theta \cos \theta \, (\omega + \Omega)^2  \nonumber \\
  & + &  \rho r \, a_\theta
        - \doverd{p}{\theta} - \rho\, \doverd{\Phi}{\theta}
   + r \ f\subscr{\theta}
      \label{eq:u_theta}
\end{eqnarray}

\noindent
d) Angular Momentum
\begin{equation}
 \doverd{\, \rho  h_t}{t} + \nabla \cdot \left( \rho\, h_t\, {\bf u} \right)
    =
         \rho  r \sin\theta \ a_\varphi 
        - \doverd{p}{\varphi} - \rho\, \doverd{\Phi}{\varphi}
   + r \sin\theta \ f\subscr{\varphi};
      \label{eq:u_phi}
\end{equation}
where we defined the {\it total} specific angular momentum
(in the inertial frame)
\begin{equation} 
    h_t  = r^2\, \sin^2 \theta \left( \omega + \Omega \right);
\end{equation}
i.e. the azimuthal velocity in the rotating frame
is given by $u_\varphi = \omega\,r\, \sin\theta$.

The Coriolis force in Eq.~(\ref{eq:u_phi}) for $u_\varphi$
(or $h_t$)
has been incorporated into the left hand side.
Thus, it is written 
in such a way as to conserve total angular momentum best.
This conservative treatment is necessary to obtain an accurate
solution of the embedded planet problem \citep{1998A&A...338L..37K}.

The function ${\bf f} = (f_r, f_\theta, f_\varphi)$ in the momentum equations
denotes the viscous
forces which are stated explicitly for the three-dimensional case in 
spherical polar coordinates in \citet{1978trs..book.....T}.
For the description of the viscosity we 
use a constant kinematic viscosity coefficient $\nu$.

\noindent
e) Energy equation (internal energy)
\begin {equation}
 \doverd{\rho c_v T}{t} + \nabla \cdot \left( \rho\, c_v T {\bf u} \right)
  =
        - p \nabla \cdot {\bf u}  +  Q^+  -  \nabla \cdot \vec{F}
      \label{eq:energy}
\end{equation}
Here $T$ denotes the gas temperature in the disc
and $c\subscr{v}$ is the specific heat at constant volume.
On the right hand side, the first term describes
compressional heating, $Q^+$ the viscous dissipation
function, and $\vec{F}$ denotes the radiative flux.
In writing eq.~(\ref{eq:energy}) we have assumed that the radiation energy density
$E=a_R T^4$ is always small in comparison to the thermal energy density $e=\rho c_v T$.
Here, $a_R$ denotes the radiation constant.
Furthermore, we utilise the one-temperature approach and write for the
radiative flux, using flux-limited diffusion (FLD)
\begin {equation}
\label{eq:raddif}
 \qquad   \vec{F}  =  -  \frac{\lambda c \, 4 a_R T^3}{\rho (\kappa + \sigma)} \, \nabla T\ ,
\end{equation}
where $c$ is the speed of light,
$\kappa$ the  Rosseland mean opacity, $\sigma$ the scattering coefficient,
and $\lambda$ the flux-limiter.
Using FLD allows us to perform stable accretion disc
models that cover several vertical scale heights. We use here the FLD approach
described in \citet{1981ApJ...248..321L} with the flux-limiter of
\citet{1989A&A...208...98K}. Its suitability for protostellar discs
has been shown in \citet{1996ApJ...461..933K,1999ApJ...518..833K}, and for
embedded planets in \citet{2006A&A...445..747K}.
In this work we use for the 
Rosseland mean opacity $\kappa(\rho, T)$ the analytical formulae
by \citet{1985prpl.conf..981L} and set the scattering coefficient $\sigma$ to zero.
To close the basic system of equations we use an ideal gas equation of
state where the gas pressure is given by  $p= R_{gas} \rho T /\mu$,
with the mean molecular weight $\mu$ and gas constant $R_{gas}$.
For a standard solar mixture
we assume here $\mu=2.35$. The speed of sound is calculated from
$c\subscr{s}=\sqrt{\gamma p / \rho}$ with the adiabatic index $\gamma =1.43$. 

\begin{figure}
 \centering
 \includegraphics[width=0.9\linwx]{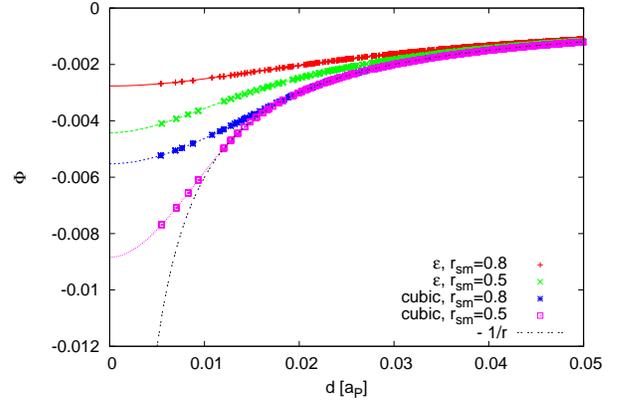}
 \caption{The gravitational potential of a $20 M_{\rm earth}$ planet with different
    smoothings applied, see Eqs.~(\ref{eq:epsilon}) and (\ref{eq:cubic}).
  The distance $d$ from the planet is given in units of $a_p$ (here $a_{Jup}$), and
   the smoothing length $r_{sm}$ in units
   of the Hill radius of the planet which refers here to $R_H = 0.0271 a_p$.
   Note that the data points are drawn directly from the used 3D computational
   grid and indicate our standard numerical resolution (see section \ref{subsec:numerics}).
   \label{fig:poti}
   }
\end{figure}

\subsection{Planetary potential}
The total potential $\Phi$ acting on the disc consists of two contributions,
one from the star $\Phi_*$, the other from the planet $\Phi_{p}$
\[
      \Phi  = \, \Phi_* + \Phi_{p} \,  = - \, \frac{G M_*}{r}
      -  \frac{G m_\mathrm{p}}{\sqrt{({\bf r}-{\bf r}_{p})^2}},
\]
where ${\bf r}_{p}$ denotes the radius vector of the planet location.
The embedded planet is modelled as a point mass that 
orbits the central star on a fixed circular orbit.
In numerical simulations the planetary
potential has to be smoothed over a few gridcells to avoid divergences.

Typically in 2D simulations the planetary potential is modelled by an
$\epsilon$-potential
\begin{equation}
\label{eq:epsilon}
   \Phi_p^{\epsilon} = - \frac{G m_p}{\sqrt{d^2 + \epsilon^2}} \, ,
\end{equation}
where we denote the distance of the disc element to the
planet with $d = |{\bf r}-{\bf r}_p|$ and $\epsilon$ is the
smoothing length.
In a 2D configuration a potential of this form is indeed very convenient,
as the smoothing takes effects of the otherwise neglected vertical
extent of the disc into account. To account for the finite disc thickness $H$
which depends on the temperature in the disc, an often used value for the
smoothing length in 2D-simulations is $\epsilon = r_{sm} = 0.6 H$.
The obtained 2D torques are then found to be in reasonable agreement 
with three-dimensional analytical estimates of the torques, that
do not include a softening length for the planet potential.

However in a 3D configuration the same approach is not necessary and would lead
to an unphysical 'spreading' of the potential over a large region.
Hence, we apply  a different type of smoothing, follow
\citet{2006A&A...445..747K} and use a {\it cubic}-potential of the form
\begin{equation}
\label{eq:cubic}
\Phi_p^{cub} =  \left\{
    \begin{array}{cc} 
   - \frac{m_p G}{d} \,  \left[ \left(\frac{d}{r_\mathrm{sm}}\right)^4
     - 2 \left(\frac{d}{r_\mathrm{sm}}\right)^3 
     + 2 \frac{d}{r_\mathrm{sm}}  \right]
     \quad &  \mbox{for} \quad  d \leq r_\mathrm{sm}  \\
   -  \frac{m_p G}{d}  \quad & \mbox{for} \quad  d > r_\mathrm{sm} 
    \end{array}
    \right.
\end{equation}
This potential is constructed in such a way as to yield for distances $d$
larger than $r_\mathrm{sm}$ the correct $1/r$ potential of the planet, and
inside that radius ($d < r_\mathrm{sm}$) the potential is smoothed with a cubic
polynomial such that at the transition radius $r_\mathrm{sm}$ the potential and
its first and second derivative agree with the analytic outside $1/r$-potential.
To illustrate the various types of smoothing, we display in Fig.~\ref{fig:poti}
the behaviour of the different forms of the planetary potential.
Clearly the $\epsilon$-potential leads
for the same values of $r_\mathrm{sm}$ to a much wider and shallower potential
than our {\it cubic}-approach. For the often used value $\epsilon = 0.6 H$ the
$\epsilon$-potential is felt way outside the Hill radius
\[
  R_{H} = a_p \left(\frac{m_p}{3 M_*}\right)^{1/3} \, ,
\]
and leads to a significant
underestimate of the potential depth already at $r_\mathrm{sm}$.
The {\it cubic}-potential (Eq.~\ref{eq:cubic}) will always be accurate down
to $d = r_\mathrm{sm}$
and is inside much deeper than the $\epsilon$-potential, and hence more accurate.
In the simulations presented below we study in detail the influence that
the potential description has on the value of the torques acting on the
planet.

We calculate the gravitational torques acting on the planet
by  integrating over the whole disc, where we apply a tapering
function to exclude the inner parts of the Hill sphere of the planet.
Specifically, we use the smooth (Fermi-type) function
\begin{equation}
\label{eq:fermi}
     f_b (d)=\left[\exp\left(-\frac{d/R_H-b}{b/10}\right)+1\right]^{-1}
\end{equation}
which increases from 0 at the planet location ($d=0$) to 1
outside $d \geq R_{H}$ with a midpoint $f_b = 1/2$ at $d = b R_{H}$,
i.e. the quantity $b$ denotes the torque-cutoff radius in units of the Hill radius.
This torque cutoff is necessary to avoid first a large, possibly very noisy
contribution from the inner parts of the Roche lobe, and second to
disregard material that is gravitationally bound to the planet.
The question of torque cutoff and exclusion of Roche lobe material becomes
very important 
when 
({\it i}) the disc self-gravity is neglected, and
({\it ii}) there exists material bound to the planet (e.g. a circumplanetary disc).
This issue should definitely be addressed in the future. 
Here, we assume a transition radius of $b = 0.8$ Hill radii \citep[see][
Fig.~2]{2008A&A...483..325C}.
For reference we quote the width of the horseshoe region which is given
for an isothermal disc approximately by \citep{2006ApJ...652..730M}
\beq
\label{eq:horseshoe}
     x_s  = 1.16 \, a_p \, \sqrt{\frac{q}{(H/r)}}
\eeq
with the mass ratio $q = m_p/M_*$ and the local relative disc thickness $H/r$.
For an adiabatic disc $H$ has to be replaced by $\gamma H$ \citep{2008ApJ...672.1054B}.
\begin{figure}[ht]
\begin{center}
\rotatebox{0}{
\resizebox{0.95\linwx}{!}{%
\includegraphics{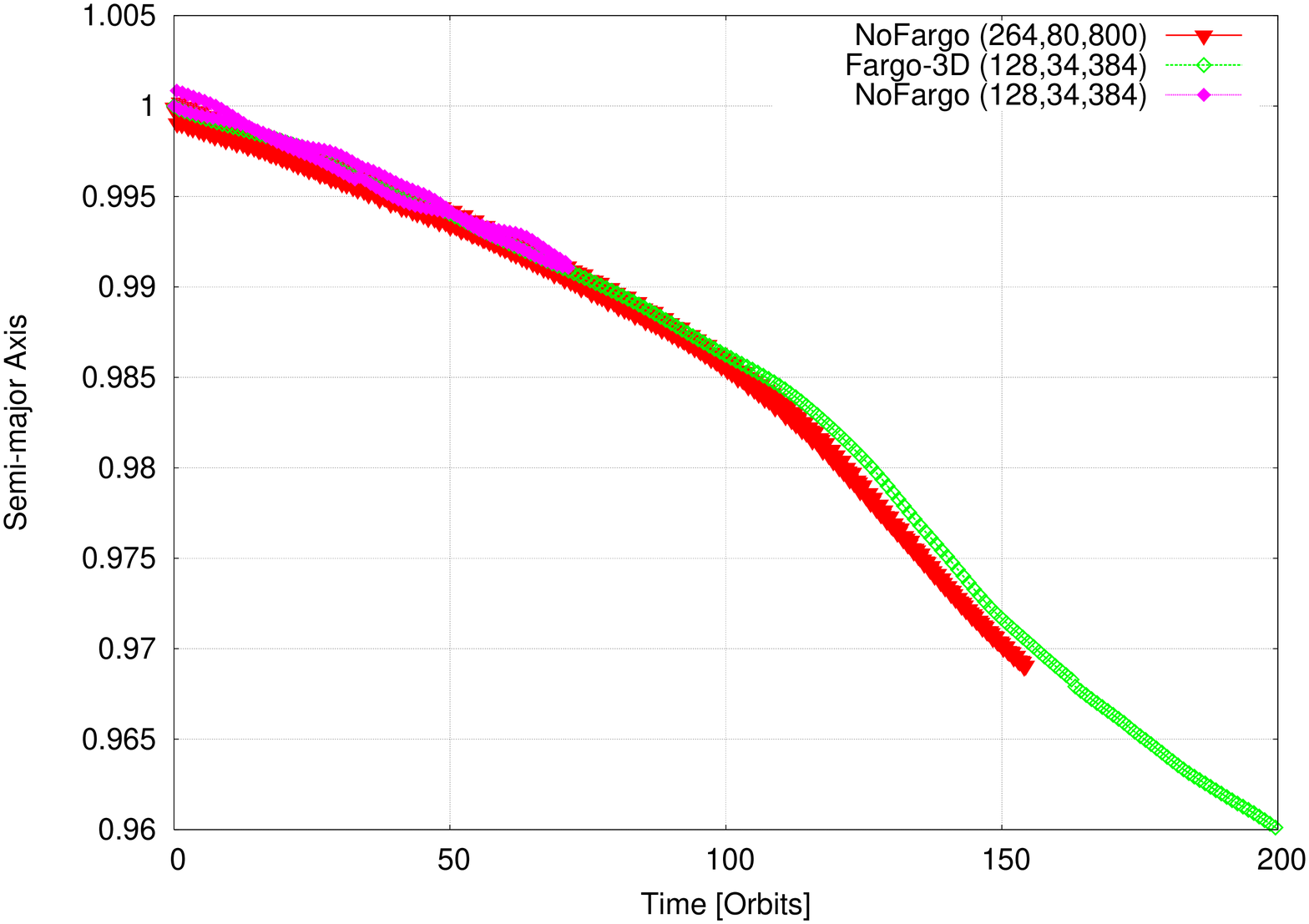}}}
\rotatebox{0}{
\resizebox{0.95\linwx}{!}{%
\includegraphics{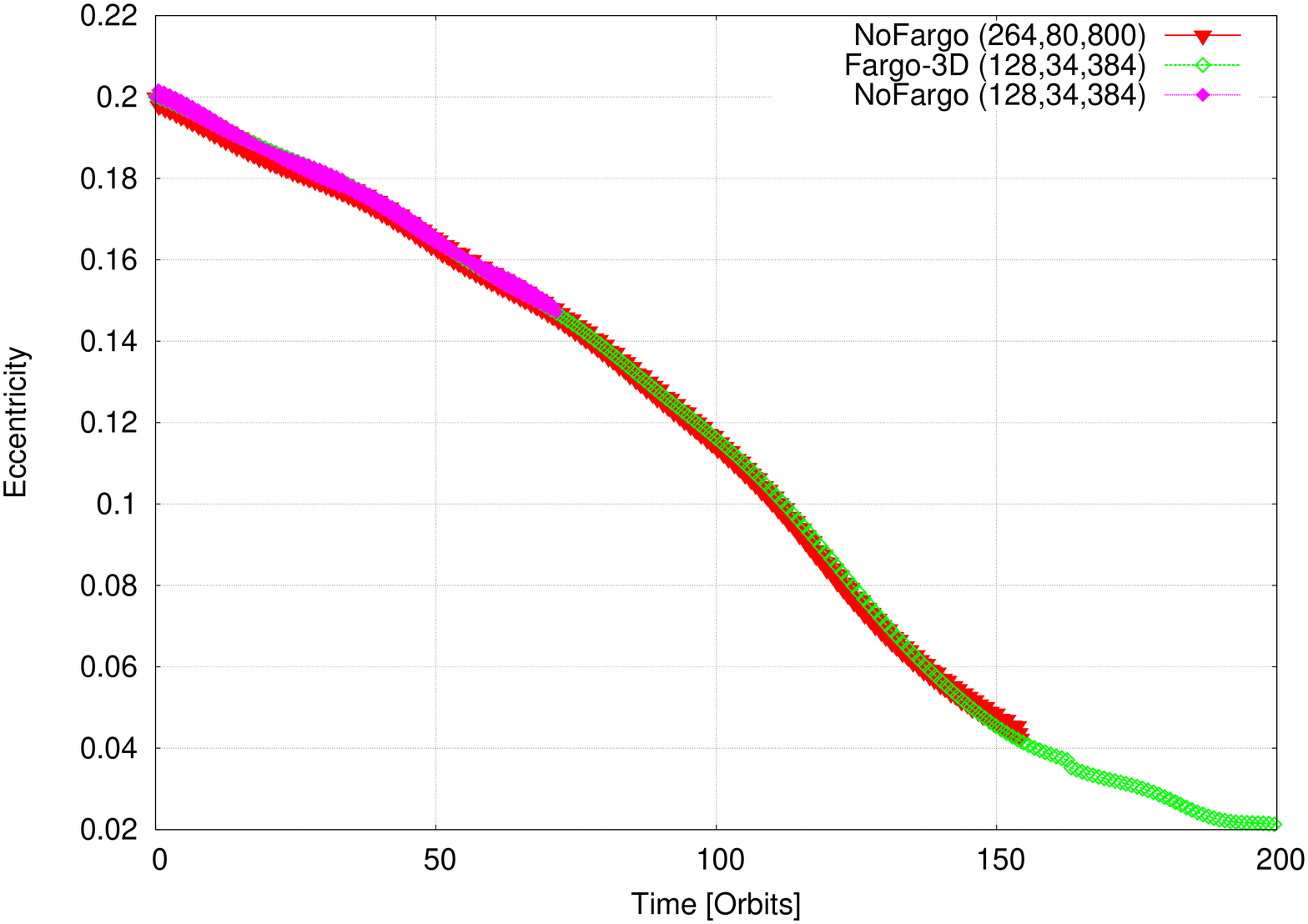}}}
\rotatebox{0}{
\resizebox{0.95\linwx}{!}{%
\includegraphics{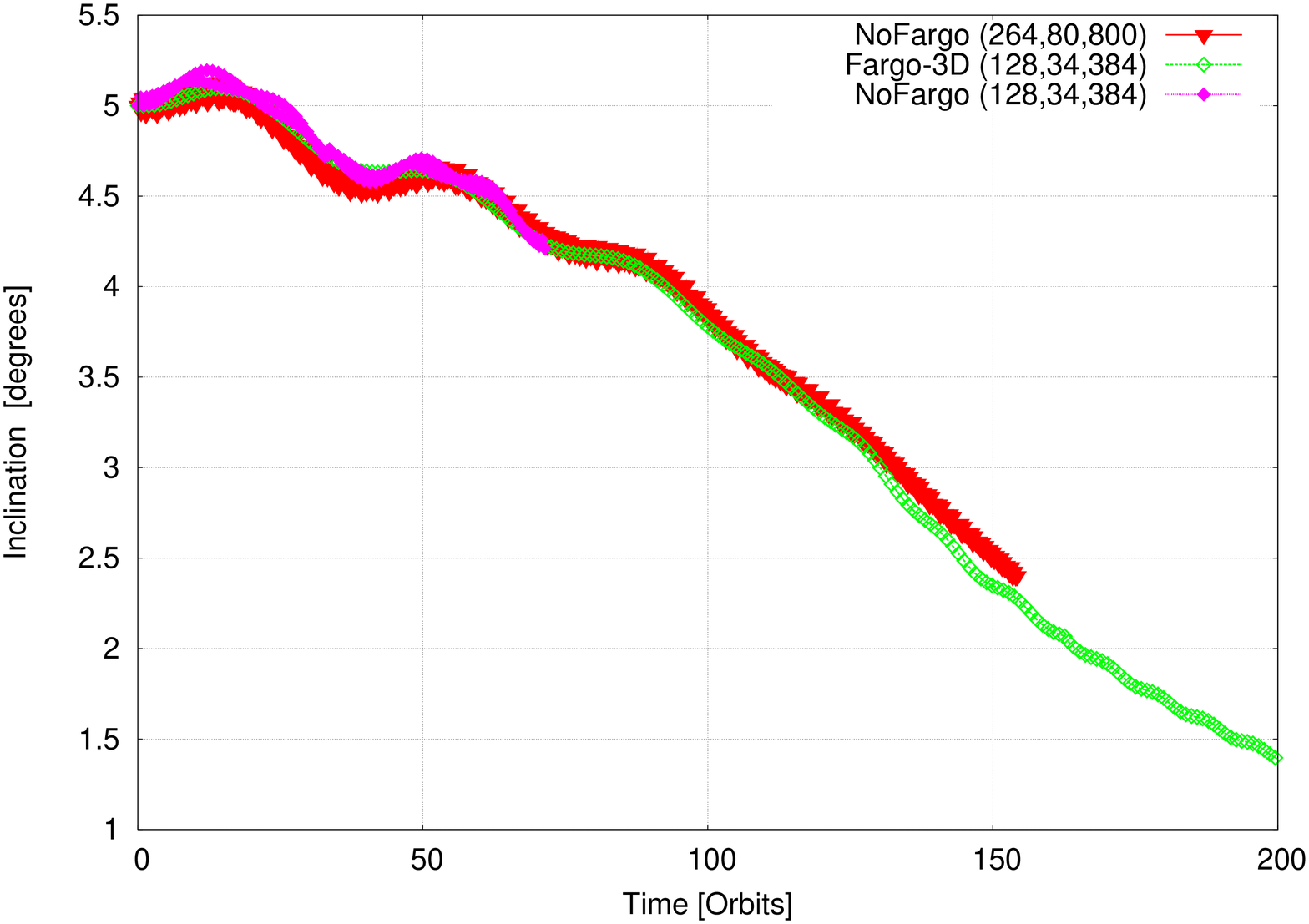}}}
\end{center}
  \caption{Evolution of semi-major axis, eccentricity and inclination as a
   function of time for a 20 $M_{earth}$ planet in a three-dimensional disc.
   Results are displayed for different numerical resolutions ($N_r, N_\theta, N_\varphi$),
   and using Fargo and Non-Fargo runs.
   }
   \label{fig:aeires3d}
\end{figure}

\subsection{Setup}
\label{subsec:setup}
The three-dimensional ($r, \theta, \varphi$) computational domain consists of a
complete annulus of the protoplanetary disc centred on the star,
extending from $r\subscr{min} = 0.4$ to $r\subscr{max} = 2.5$ in units
of $r_0 = a_{Jup} = 5.2$AU. In the vertical direction the annulus extends
from the disc's midplane
(at $\theta = 90^\circ$) to about $7^\circ$ (or $\theta = 83^\circ$) above
the midplane. In case of an inclined planet the domain has to be extended
and cover the upper and lower half of the disc.
The mass of the central star is one solar
mass $M_* = M_\odot$, and the total disc mass inside $[r\subscr{min},r\subscr{max}]$
is $M_{disc} = 0.01 M_\odot$. For the present study, we use a constant
kinematic viscosity coefficient with a value of $\nu = 10^{15}$\,cm$^2$/s,
a value that relates to an equivalent $\alpha = 0.004$ at $r_0$  for a
disc aspect ratio of $H/r = 0.05$, where $\nu = \alpha H^2 \Omega_K$.
In standard dimensionless units we have $\nu = 10^{-5}$.

The models are initialised with a locally isothermal configuration
where the temperature is constant on cylinders and has the
profile $T(s) \propto s^{-1}$, where $s$ is related to $r$ through
$s = r \sin \theta$.
This yields a constant ratio of the disc's vertical height $H$ to the
radius $s$. The initial vertical density stratification is approximately given by
a Gaussian:
\begin{equation}
  \rho(r,\theta)= \rho_0 (r) \, \exp \left[ - \frac{(\pi/2 - \theta)^2 \, r^2}{H^2} \right].
\end{equation} 
Here, the density in the midplane is $\rho_0 (r) \propto r^{-1.5}$ which
leads to a $\Sigma(r) \propto \, r^{-1/2}$ profile of the vertically integrated
surface density.
The vertical and radial velocities, $u_\theta$ and $u_r$, are initialised to zero.
The initial azimuthal velocity $u_\varphi$ is
given by the equilibrium of gravity, centrifugal acceleration and the
radial pressure gradient.
In case of purely isothermal calculations this setup is equal to the equilibrium
configuration (in the case of closed radial boundaries).
For fully radiative simulations the model is first run in a 2D axisymmetric
mode to obtain a new self-consistent equilibrium where viscous heating
balances radiative transport/cooling from the surfaces 
(see Sect.~\ref{subsec:initial} below).
This initialisation through an axisymmetric  2D phase (in the $r-\theta$ plane) reduces the required
computational effort substantially, as
the evolution from the initial isothermal state towards the radiative equilibrium
takes about 100 orbits, if the disc is started with an isothermal equilibrium
having constant $H/r$.
After reaching the equilibrium between
viscous heating and radiative transport/cooling, we extend this model to
a full 3D simulation, by expanding the grid into the $\phi$-direction,
and the planet is embedded.

\subsection{Numerics}
\label{subsec:numerics}
We adopt a coordinate system, which rotates at the orbital
frequency of the planet. For our standard cases, we use an
equidistant grid in $r,\theta, \varphi$ with a resolution of 
$(N_r,N_\theta, N_\varphi) = (266,32,768)$.
To minimise disturbances (wave reflections) from the radial boundaries,
we impose, at $r\subscr{min}$ and $r\subscr{max}$, damping boundary
conditions where all three velocity components are relaxed towards their
initial state on a timescale of approximately the local orbital
period.  The radial velocities at the inner and outer radius vanish.
The angular velocity is relaxed towards
the Keplerian values.  For the density and temperature, we apply
reflective radial boundary conditions. In the azimuthal direction,
periodic boundary conditions are imposed for all variables.
In the vertical direction we apply outflow boundary conditions.
The boundary conditions do not allow for mass accretion through the disc,
such that the total disc mass remains nearly constant during the time evolution,
despite a possible small change due to little outflow through the
vertical boundaries and the used density floor (see below).

The numerical details of the used finite volume code ({\tt NIRVANA})
relevant for these planet disc simulations were described in
\citet{2001ApJ...547..457K} and \citet{2003ApJ...586..540D}. In the latter paper the
usage of the nested grid-technique is described in more detail as well.
The original version of the {\tt NIRVANA} code, on which our programme
is based upon, has been developed by \citet{1997CoPhC.101...54Z}. 
The empowerment with {\tt FARGO} is based on the original work
by \citet{2000A&AS..141..165M}. Our implementation appears to be the first
inclusion of the {\tt FARGO}-algorithm in a 3D spherical coordinate system.
More details about the implementation are given in the appendix.
The basic algorithm of the newly implemented radiation part in the
energy equation (\ref{eq:energy}) is presented in the appendix as well. 
To avoid possible time step limitations this part is always solved implicitly.

\section{Test calculations}
\label{sec:tests}
\subsection{The {\tt FARGO}-algorithm}
To test the 3D implementation of the {\tt FARGO}-algorithm in our
{\tt NIRVANA}-code we
have run several models with planets on circular, elliptic and inclined
orbits with and without the {\tt FARGO}-method applied.
Here, we follow closely the models presented in \citet{2007A&A...473..329C}
and consider moving planets in 3D discs. As the tests are dynamically already
complicated we use here only the isothermal setup.
The different setups gave very similar results in all cases, and 
we present results for one combined case of a 20~$M_{\rm earth}$ planet
embedded in a locally isothermal disc with an initial non-zero eccentricity
($e=0.2$) and non-zero inclination ($i=5^\circ$).
All physical parameters of this run are identical to those described
in  \citet{2007A&A...473..329C}, and we compare our results to the last models
presented in that paper (their Fig.~16).
The outcome of this comparison is shown in Fig.~\ref{fig:aeires3d}, where we display
the results of a standard non-Fargo run with the resolution
$(N_r,N_\theta, N_\phi)= (264,80,800)$ 
with the data taken from \citet{2007A&A...473..329C} (where a different 
code has been used)
to two runs having a lower resolution of $(N_r,N_\theta, N_\varphi) = (128,34,384)$,
one with {\tt FARGO} and the other one without.
We can see that all 3 models (obtained with two different codes, methods and resolutions)
yield very similar results. 
The scatter of the data points is slightly smaller in the {\tt FARGO}-run.

\begin{figure}
 \centering
 \includegraphics[width=0.49\linwx]{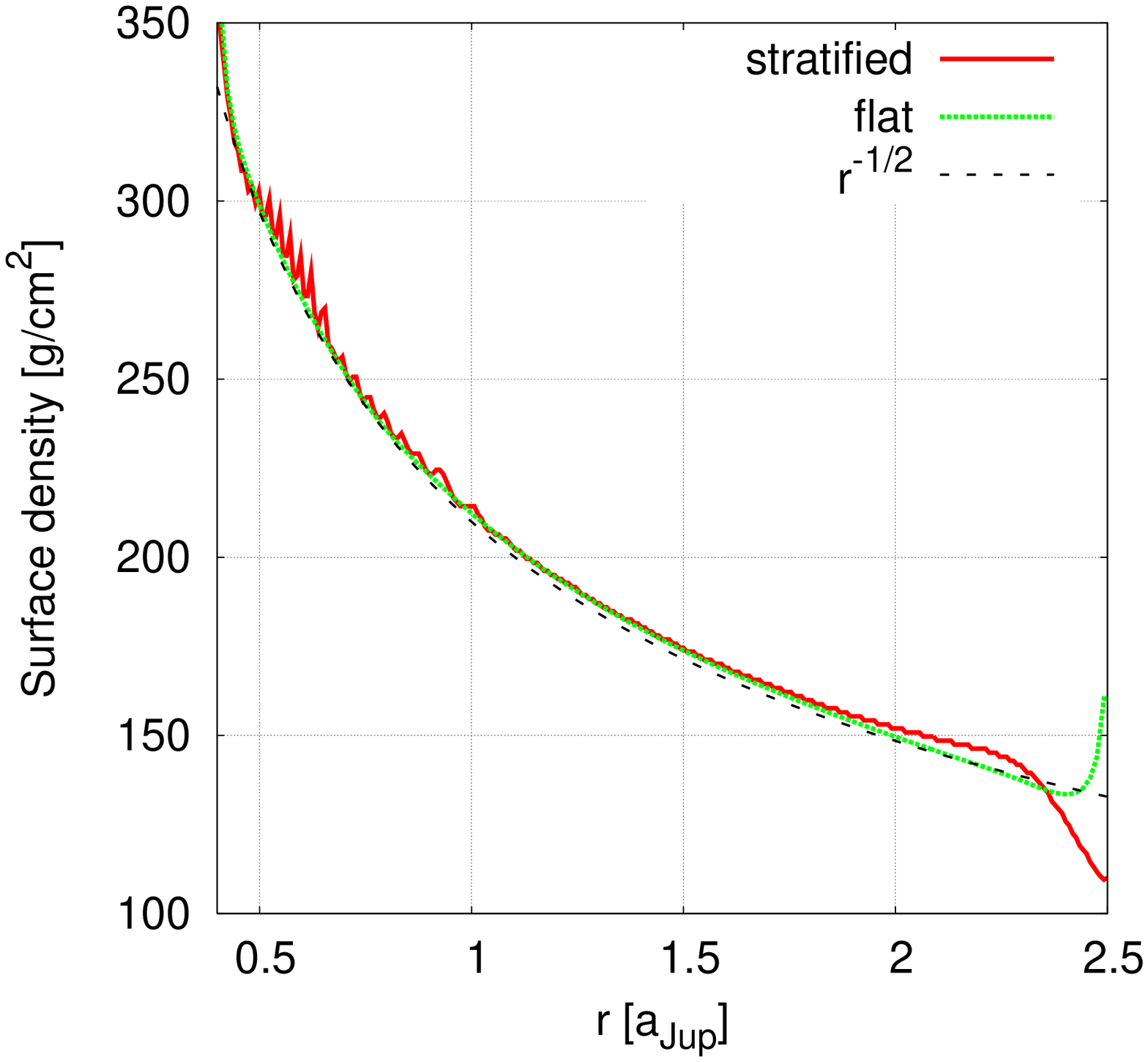}
 \includegraphics[width=0.49\linwx]{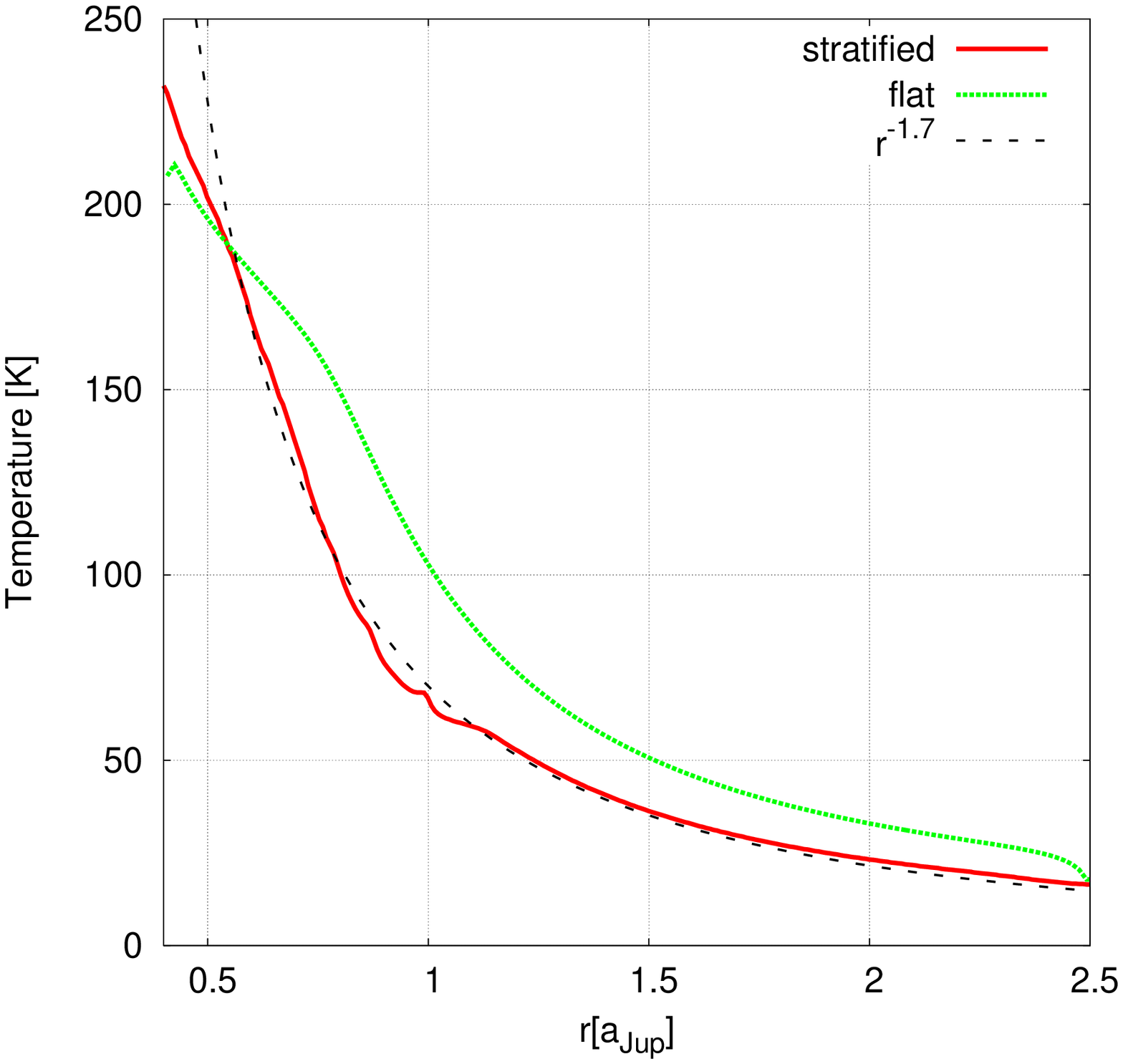}
 \caption{Radial stratification of surface density (left) and the midplane temperature (right) 
   in the disc. This dashed lines represent simple approximations to the 3D stratified results.
   The solid (green) curves labelled 'flat' refer to corresponding results for a
   vertically integrated flat 2D disc using the same input physics.
   \label{fig:radial} 
   }
\end{figure}

\begin{figure}
 \centering
 \includegraphics[width=0.49\linwx]{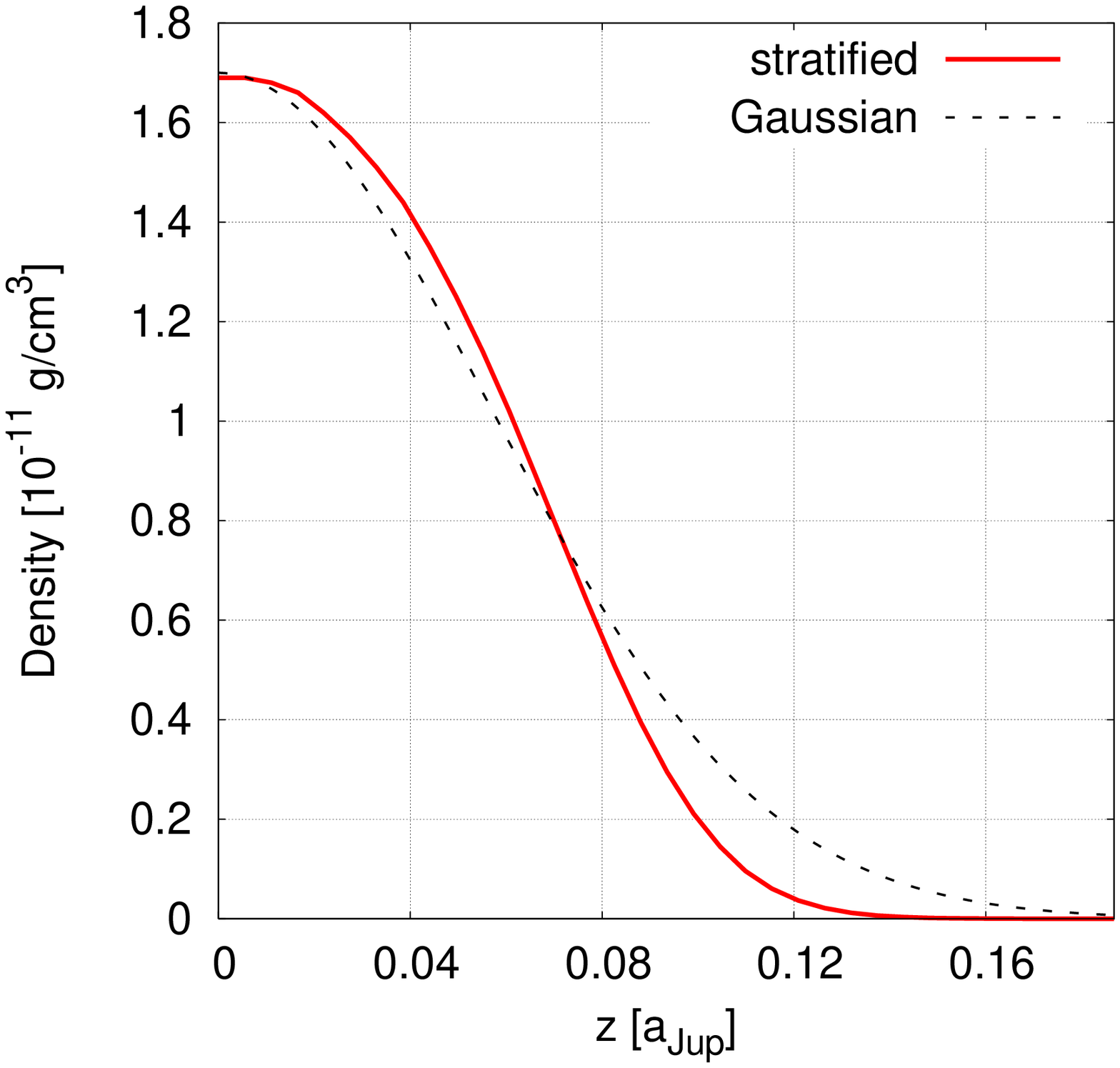}
 \includegraphics[width=0.49\linwx]{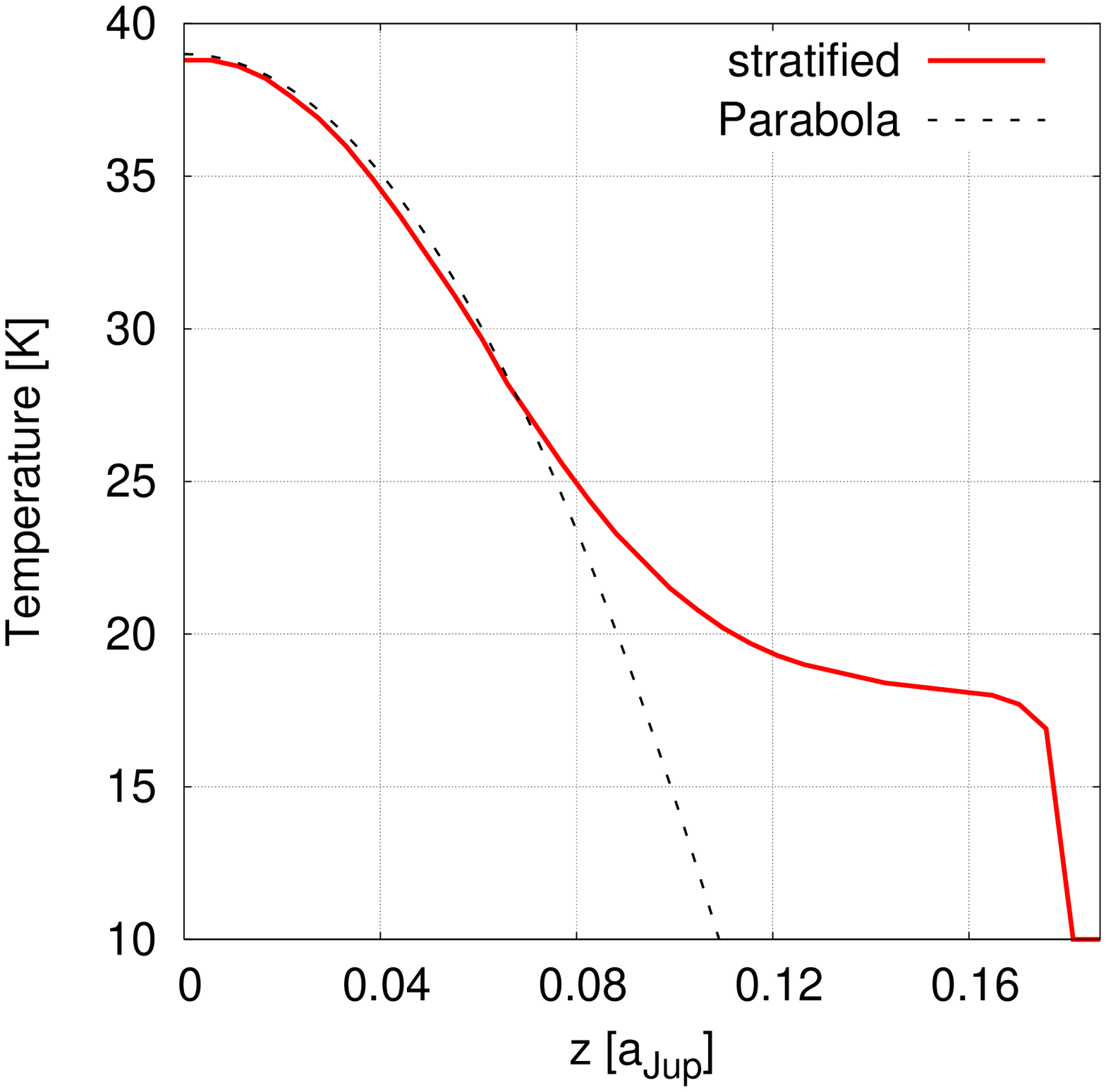}
 \caption{Vertical stratification of density (left) and temperature (right) at a radius
   of $r=1.44$. Thin dashed lines just represent simple functional relations.
   \label{fig:vertical}    
   }
\end{figure}

\subsection{The radiation algorithm}
To obtain an independent  test of the newly implemented radiation transport
module in our {\tt NIRVANA}-code we performed a run with the standard setup
as described above but with no embedded planet. Hence, this setup
refers to an axisymmetric disc with internal heating and radiative
cooling. For a fixed, closed computational domain it is only the
total mass enclosed that determines the final equilibrium state of the system,
once the physics (viscosity, opacity, and equation of state) have been prescribed.
The radial dependence of the vertically integrated surface density and the 
midplane temperature are
displayed in Fig.~\ref{fig:radial}, and the corresponding vertical
profiles at a radius of $r=1.44$ in  Fig.~\ref{fig:vertical}.
First of all, these new results obtained with {\tt NIRVANA}
agree very well with those obtained with the completely
independent 2D-code {\tt RH2D} used in the $r-\theta$ mode,
such as presented for example in \citet{1993ApJ...416..679K},
which are not shown in the figures, however. 
As the final configuration of the system is given by the equilibrium of
internal (viscous) dissipation and radiative transport, this test demonstrates
the consistency of our implementations.

To relate our 3D results to previous radiative 2D runs which use
vertically integrated quantities, and hence can only use an approximative
energy transport and cooling \citep{2008A&A...487L...9K}, we
compare in Fig.~\ref{fig:radial} the results obtained with the two methods.
Both models are constructed for the same disc mass and identical physics.
The label 'flat' in the Figure refers to the flat 2D case (obtained with
{\tt RH2D}, see \citet{2008A&A...487L...9K}) 
and the 'stratified' label to our new 3D implementation 
presented here. The left graph displays the vertically integrated
surface density distribution, 
here $\Sigma(r) = f_\Sigma \int_{\theta_{max}}^{\theta_{min}} \rho r \sin(\theta) d\theta$,
with $f_\Sigma = 1$ for two-sided and 2 for one-sided discs.
Our result is well represented by a $\Sigma \propto r^{-1/2}$  profile as expected
for a closed domain and constant viscosity. Interesting is the irregular
structure at radii smaller than $r\approx 1.0$ in the full 3D stratified case,
and we point out that these refer to the onset of convection inside that
radius. To model convection is of course not possible 
in a flat 2D approach.
The temperature distribution for the full 3D case follows approximately a
$T \propto r^{-1.7}$ profile. Here, the approximate flat-disc model
leads to midplane temperatures that are about 40\% higher for the bulk part of the
domain than in the true 3D case. Possibly a refined modelling of the vertical 
averaging procedure and the radiative losses
in the flat 2D case could improve the agreement here, but in the presence of
convection we may expect differences in any case.

In Fig.~\ref{fig:vertical} we display the vertical stratification of the disc at a specific
radius in the middle of the computational domain at $r=1.44$. Two simple
approximations are over-plotted as dashed lines. 
Note, that in these plots the stratification
is plotted along the $r=const.$ lines which deviates for thin discs 
only slightly from $z = r \cos \theta$. 
Taking $z_0=0.08$, the Gaussian curve 
for the density refers to $\rho_0 \exp\left[-(z/z_0)^2\right]$ and the temperature fit
to $T(z) = T_0 [ 1 - 0.4 \, (z/z_0)^2]$. 
These simple formulae are intended to guide the eye rather
than meant to model exactly the structure at this radius which depends on the
used opacities. Given the simplicity of these, it is interesting that they
approximate the true solution reasonably well within one scale height.


\subsection{Density Floor}

By expanding the computed area in the $\theta$-direction beyond the $90^\circ$ to $83^\circ$
region of our standard model,
the code would have to cover several orders of magnitude in the density.
Thus, many more grid cells would be required to resolve the physical quantities. 
In order to avoid this and save computation time, we apply a minimum density
function (floor) for the low-density
regions high above the equatorial plane of the disc. It reads
\begin{equation}
	\rho = \left\lbrace \begin{array}{cc}
		\rho &  \quad \mbox{for} \quad \rho > \rho_{min} \\
		\rho_{min} & \quad \mbox{for} \quad \rho \le \rho_{min}
	       \end{array} \right. \ .
\end{equation}

\begin{figure}
 \centering
 \includegraphics[width=1.0\linwx]{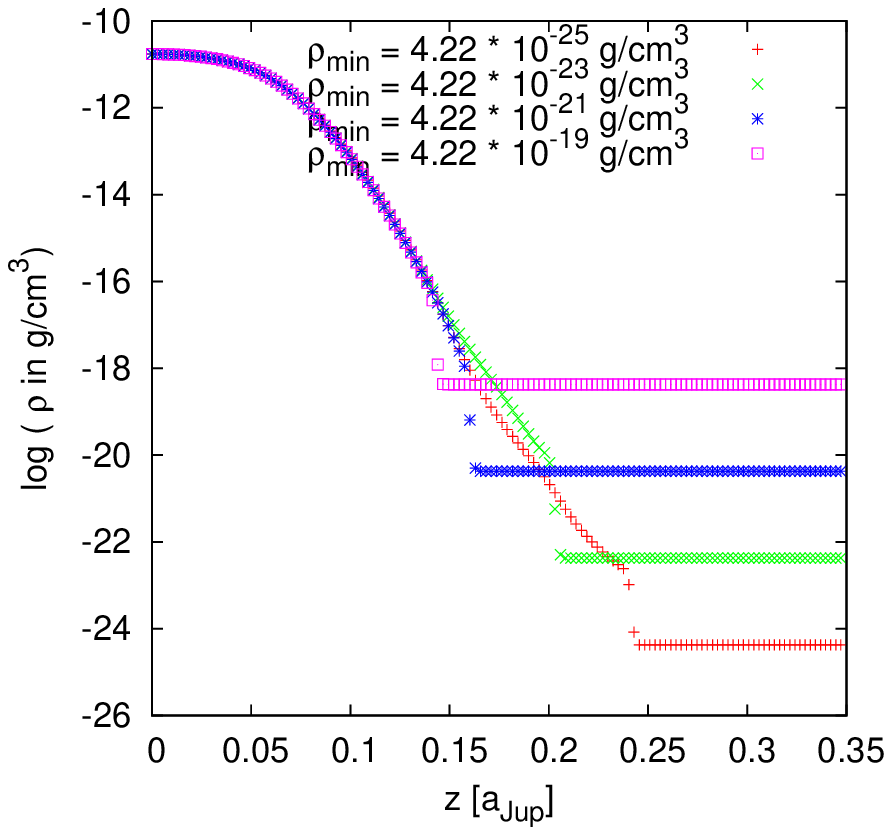}
 \includegraphics[width=1.0\linwx]{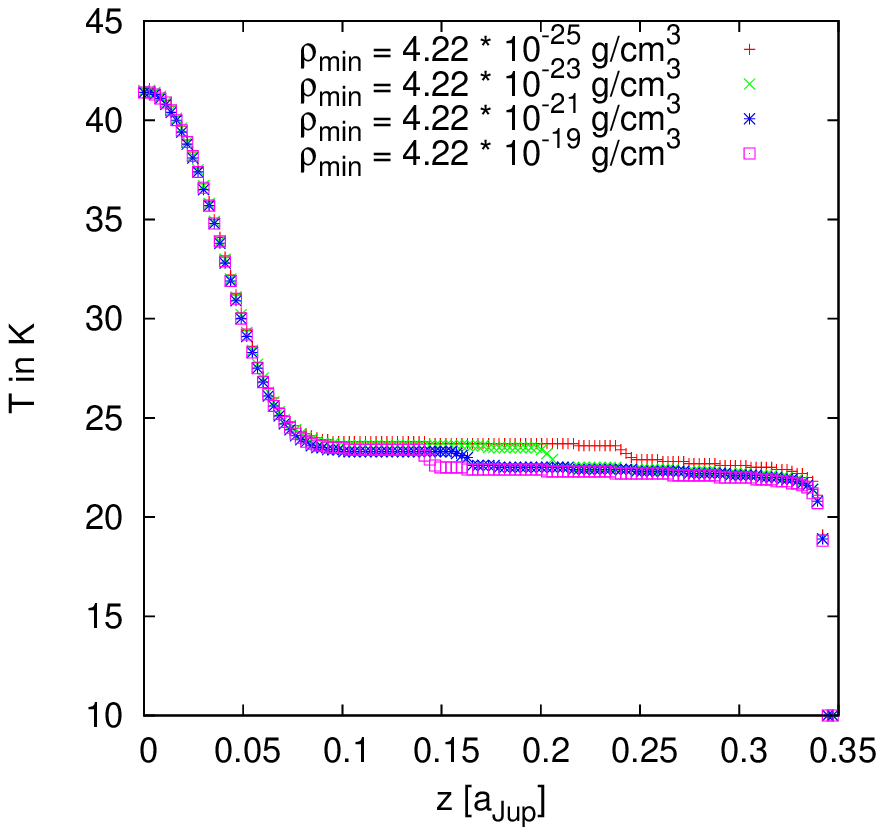}
 \caption{Vertical stratification of density in logarithmic scale (top)
 and temperature (bottom) at a radius
   of $r=1.00$ for a simulation covering the $\theta = 90^\circ$ to $70^\circ$ region.
 The optical depth $\tau=1.0$ is reached at about $z=0.055$.
   \label{fig:verticalRHOMIN}
   }
\end{figure}

Of course, applying a density floor like this will create mass inside the computed
domain. The density floor $\rho_{min}$ has now to be chosen such that: firstly the computation
is not handicapped by too small values and secondly the inner (optically thick)
parts of the disc are not influenced.
To test the sensitivity of the disc structure against the density floor
we performed a series of test calculations,
and show the results of simulations with different $\rho_{min}$ in
Fig.~\ref{fig:verticalRHOMIN}. These runs cover $\theta = 90^\circ$ to $70^\circ$,
a range about 3 times as large as before.
The density and temperature profiles in these simulations do not differ for
the regions near the equatorial plane, because the density is too high for the
minimum density to take effect. Indeed, all curves are nearly indistinguishable 
in the region for optical depths larger than $\tau =1.0$, with
\beq
 \tau(z) = \int_z^\infty \rho \kappa dz.
\eeq
Please note, that in the plot we do display the results along lines of constant (spherical)
radius.
Moving further away from the equatorial plane,
one can see in the density profile the different minimum densities, but in the
temperature profile there is hardly any difference at all. Above a certain distance from
the equatorial plane the temperature remains constant. The little fluctuations
visible in the profile are due to oscillations in the temperature for the low mass regions.

By applying a minimum density the code is capable of resolving large distances
above the equatorial plane with a reasonable number of grid cells.
Also note that it is not necessary to use a minimum density for calculations
covering only the $\theta = 90^\circ$ to $83^\circ$ regions, as the density is
always high enough.

\section{Models with an embedded planet}
\label{sec:embedded}
For all the models with embedded planets we use our standard disc setup as
described in section~\ref{subsec:setup} with the corresponding boundary conditions
in section~\ref{subsec:numerics}.
Here, we briefly summarise some important parameter of the setup.
The three-dimensional ($r, \theta, \varphi$) structure of the disc extends form
$r\subscr{min} = 0.4$ to $r\subscr{max} = 2.5$
in units of $r_0 = a_{Jup} = 5.2$AU. In the vertical direction the annulus
extends from the disc's midplane (at $\theta = 90^\circ$) to about $7^\circ$
(or $\theta = 83^\circ$) above the midplane.
For our chosen grid size of $(N_r, N_\varphi, N_\theta) = 266, 32, 768)$ this
refers to linear grid resolution of $\Delta \approx 0.008$ at the location of
the planet, which corresponds to $3.3$ gridcells per Hill radius, and to
about 5 gridcells per horseshoe half-width for a $20 M_{earth}$ planet
in a disc with $H/r =0.05$. 
In this configuration the planet is located exactly at the corner of a gridcell.
In the fully radiative disc, the temperature at the disc surface is kept at the
fixed ambient temperature of 10 K. 
This simple 'low-temperature' boundary condition ensures that all the internally generated
energy is liberated freely at the disc's surface. It is only suitable for optically thin
boundaries and does not influence the inner parts of the optically thick disc
(see Fig.~\ref{fig:verticalRHOMIN}).
The disc has a mass of $0.01 M_\odot$, and an aspect ratio $H/r = 0.05$ in the beginning.

\subsection{Initial setup}
\label{subsec:initial}
Before placing the planet into the 3D disc we have to bring it first into a
radiative equilibrium state such that our results are not corrupted by initial transients.
As described above this initial equilibration is performed in an axisymmetric 2D setup that
is then expanded to full 3D.
Tests with our code have shown that we reach the 3D equilibrium state
(a constant torque) in a calculation with embedded planets
about $50 \%$ faster when starting first with the 2D radiative equilibrium disc.

\begin{figure}
 \centering
 \includegraphics[width=0.98\linwx]{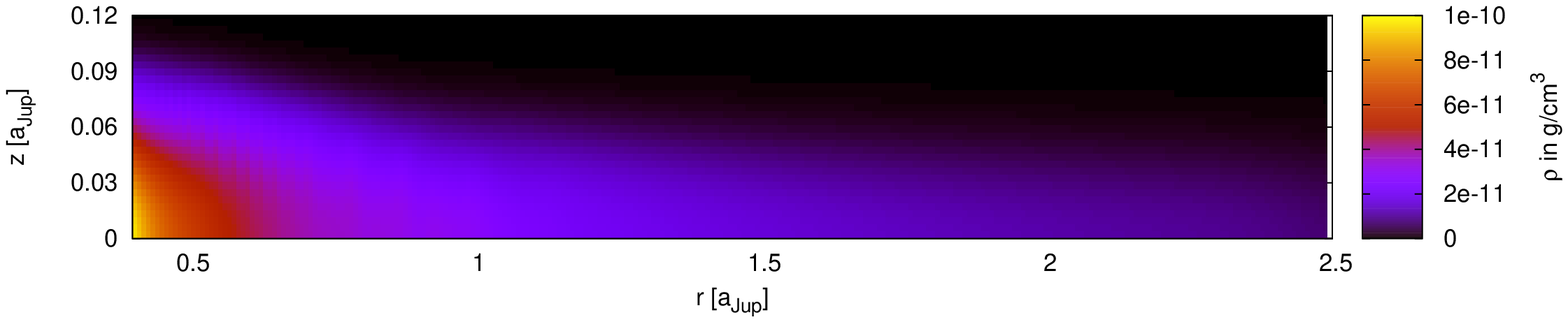} \\
 \includegraphics[width=0.98\linwx]{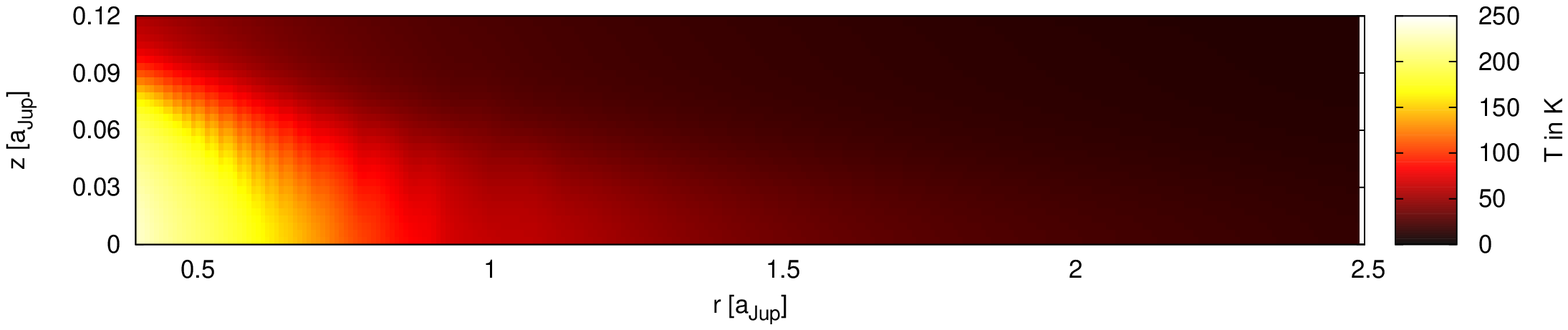}
 \caption{Density (top) and Temperature (bottom) for a fully radiative model in a
  2D axisymmetric simulation
   \label{fig:2DDisc}
   }
\end{figure}

\begin{figure}
 \centering
 \includegraphics[width=0.9\linwx]{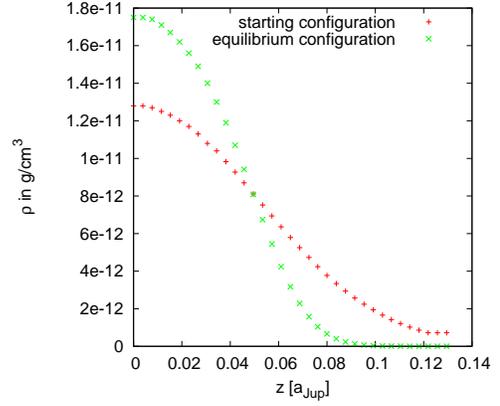}
 \caption{Vertical density distribution at $r=1.4$ for a fully radiative model
  in a 2D axisymmetric
 simulation. \textcolor{red}{'+'}: isothermal starting configuration,
 \textcolor{green}{'x'}: relaxed radiation equilibrium configuration
   \label{fig:2DDisc1DRho}
   }
\end{figure}

In Fig.~\ref{fig:2DDisc} the 2D density and temperature distributions for
such an equilibrated disc are displayed.
In the equilibrium state of the fully radiative model the disc is much thinner
in comparison to the isothermal starting case, see Fig.~\ref{fig:2DDisc1DRho}.
Consequently, the density is increased in the equatorial plane, leaving
the areas high above and below the disc with less material.
Apparently, for this disc mass and the chosen values of viscosity and
opacity, the balance of viscous heating and radiative cooling
reduces the aspect ratio of the disc from initially $0.05$ to about $0.037$
in the radiative case. Had we started with an initially thinner disc, the difference
would of course not be that pronounced.

After successfully completing the equilibration we now embed a 20 $M_{\rm earth}$ planet
into the disc. 
The planet is held on a fixed orbit and we calculate the torques acting on it
through integrating over the whole disc taking into account the above tapering
function with a cutoff $r_{torq} = 0.8 R_{H}$, which refers to $b=0.8$ in Eq.~\ref{eq:fermi}.
In addition to this value of the torque cutoff he have tested how the obtained
total torque changes when using $b=0.6$. For our standard $20 M_{\rm earth}$ planet
presented in the following we found that for the isothermal cases
the results change by about 10\% and in the radiative case by about 30\%,
which can be considered as a rough estimate of the numerical uncertainties of the
results. The deviation is larger in the radiative situation because
in this case important (corotation) contributions to the total torque originate from a region very
close to the planet, which is influenced stronger by the applied torque cutoff. 
Here, cancellation effects caused by adding the negative Lindblad and the positive
corotation torque may explain part of the larger relative uncertainty in the radiative case.
We note, that our applied
torque cutoff is not hard but refers to the smooth function (\ref{eq:fermi}).
Keeping in mind that there are only 3.3 gridcells per Hill radius, smaller
values for $b$ are not useful.

\subsection{Isothermal discs}

\begin{figure}
 \centering
 \includegraphics[width=0.9\linwx]{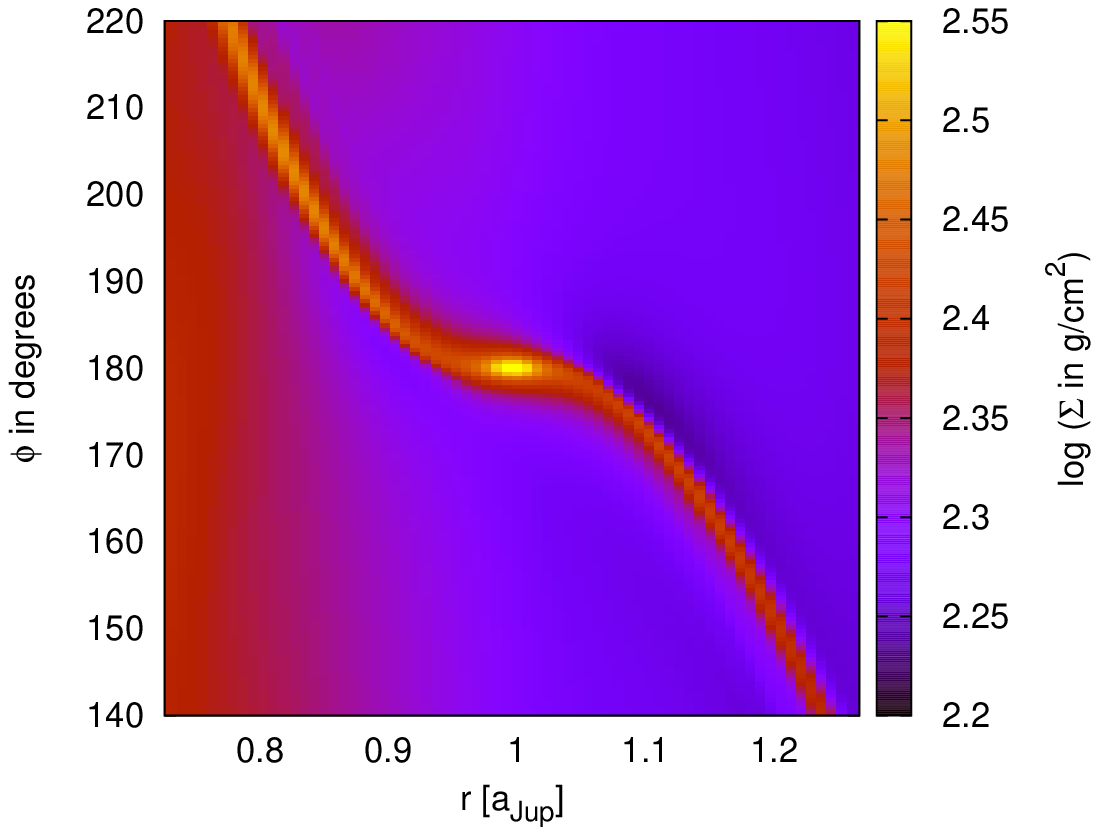}
 \includegraphics[width=0.9\linwx]{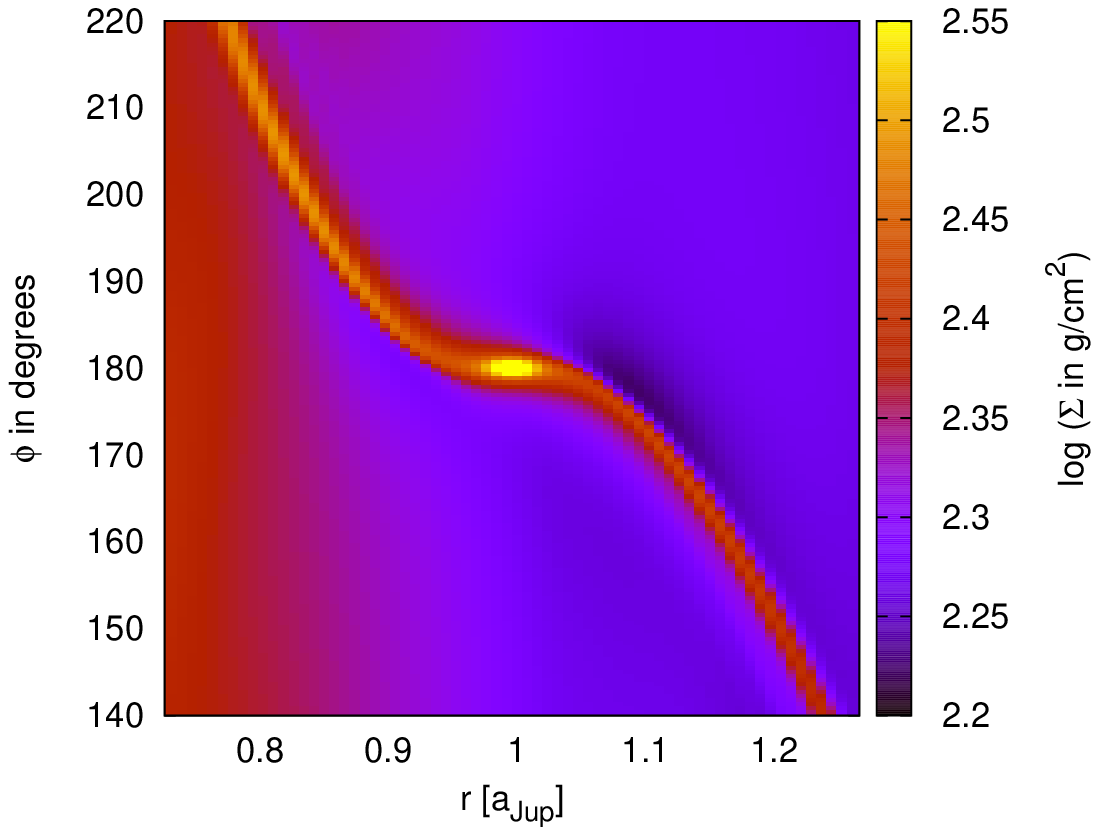}
 \caption{Surface density distribution for isothermal simulations with $H/r=0.05$ at 100
 planetary orbits. 
 Displayed are results for the shallowest and deepest potential.
  \textcolor{red}{Top}: $\epsilon$-potential with $r_{sm}=0.8$, 
  \textcolor{red}{Bottom}: cubic with $r_{sm}=0.5$.
   \label{fig:Iso2dDensity}
   }
\end{figure}

\begin{figure}
 \centering
 \includegraphics[width=0.9\linwx]{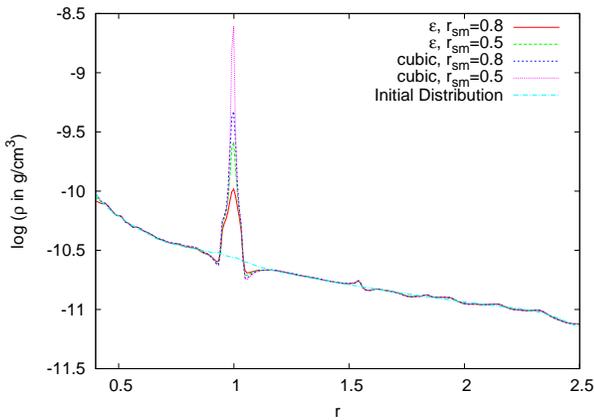}
 \caption{Radial density distribution  in the equator along
 ($\rho(r, \varphi=\pi, \theta=\pi/2)$), i.e. along a ray through the location
  of the planet for all 4 planetary potentials used.
   \label{fig:Iso1dDensity}
   }
\end{figure}

Due to the applied smoothing, we expect the 
planetary potential to modify the density structure of the disc near the
planet and subsequently change the torques acting on the planet.
First, we investigate the influence of the planetary potentials
(see Fig.~\ref{fig:poti}) on the disc and torques in the isothermal regime.
The 2D surface density distribution in the disc's midplane at 100 planetary orbits corresponding to
our two extreme planetary potentials (the shallowest and the deepest) 
is displayed in Fig.~\ref{fig:Iso2dDensity}, where we used a cutoff for the maximum displayed
density to make both cases comparable.
As expected, a deeper planetary potential results in a higher density
concentration inside the planetary Roche lobe and to a slightly reduced density
in the immediate surroundings.
This accumulation of mass near the planet for
deeper potentials is illustrated in more detail Fig.~\ref{fig:Iso1dDensity}. 
For our deepest $r_{sm} = 0.5$ cubic potential the maximum density inside
the planet's Roche-lobe is over 
an order of magnitude larger than in the shallowest $r_{sm}=0.8$ $\epsilon$-potential.

\begin{figure}
 \centering
 \includegraphics[width=0.9\linwx]{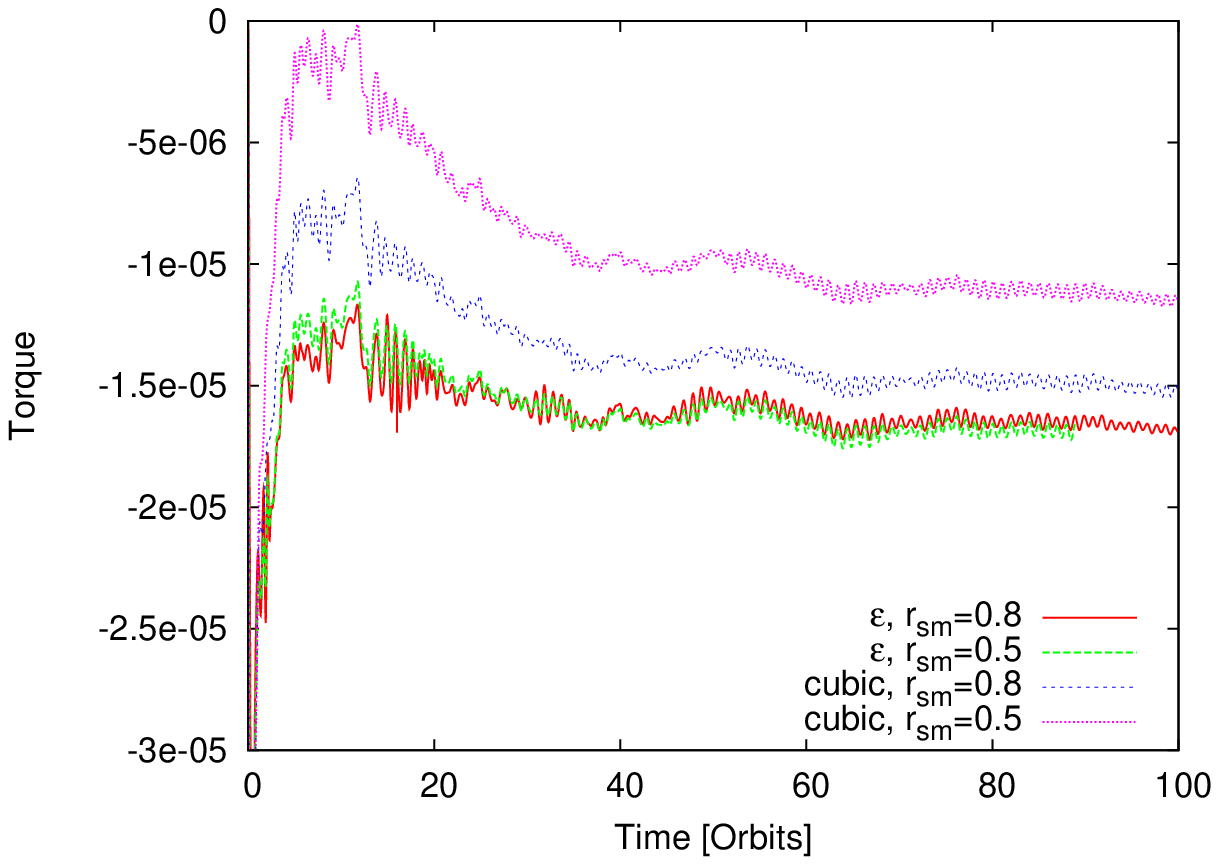} \\
 \includegraphics[width=0.9\linwx]{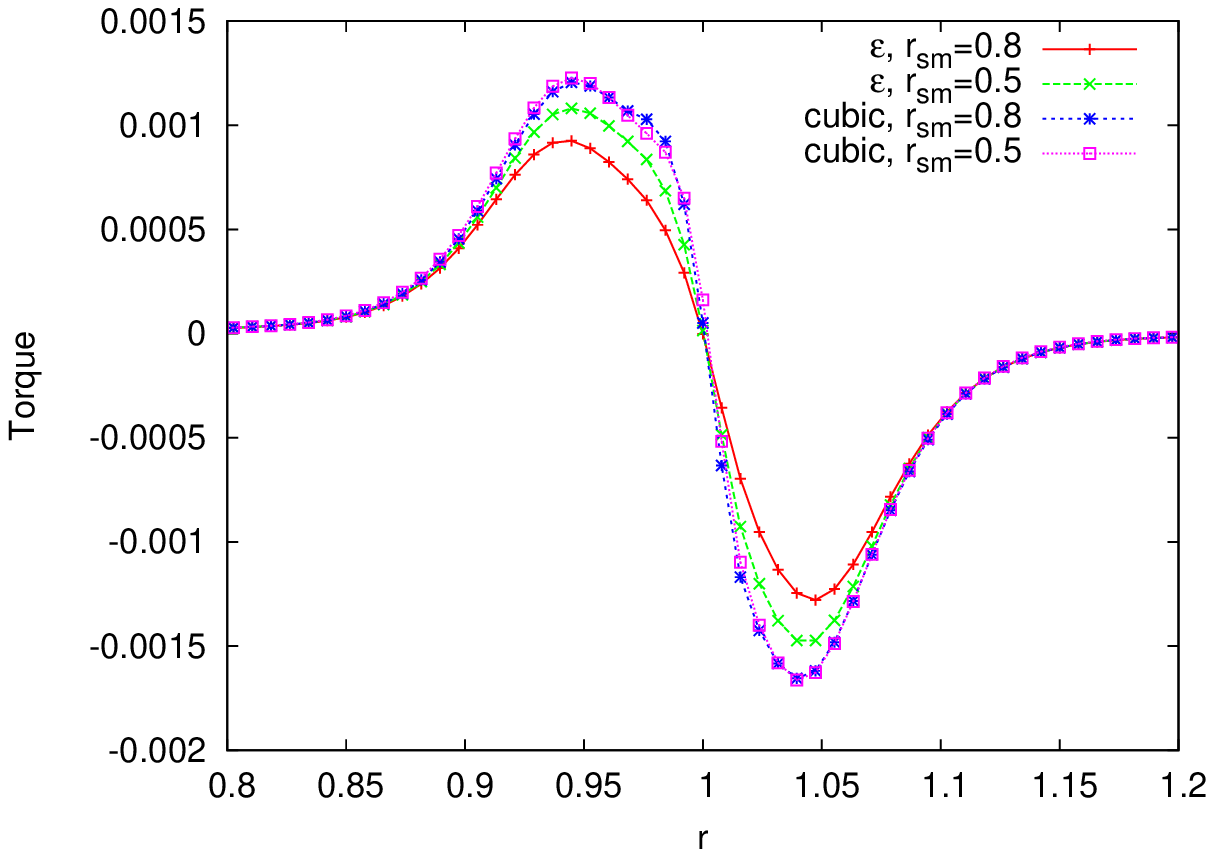}
 \caption{Specific torque (in units of $a_p^2 \Omega_p^2$)
 acting on the planet using 4 different smoothings for the
 planetary potential in the isothermal case with $H/r = 0.05$.
  {\bf Top}: Evolution of total torque with time.
  {\bf Bottom}: Radial variation of the specific torque density $\Gamma(r)$ at $t=80$ orbits.
   \label{fig:TorqueIso}
   }
\end{figure}

In Fig.~\ref{fig:TorqueIso} we show the specific torques acting onto the planet using
different potentials for the case of $H/r = 0.05$. The total torque is continuously
monitored and plotted versus time in the upper panel.
The radial torque density $\Gamma(r)$ for the same models is displayed in the lower panel.
Here, $\Gamma(r)$ is defined such that the total torque $T^{tot}$ acting on the planet
is given by
\beq
       T^{tot} = \int_{r_{min}}^{r_{max}} \, \Gamma(r) \, dr.
\eeq
The time evolution of the total torque displays a characteristic behaviour.
Starting from the axisymmetric case, a first intermediate plateau  
is reached at early times between $t\approx 5 - 10$,
after which the torques settle on longer timescales towards their final equilibria.
The initial plateaus correspond to the values of the torques shortly after the
disc material has started its horseshoe-type motion in the co-orbital region.
The level of this so-called unsaturated torque depends on the local disc properties
and on the applied smoothing of the potential, as indicated clearly in 
the top panel of Fig.~\ref{fig:TorqueIso}.   
In the following evolution, the material in this horseshoe region
will be mixed thoroughly, the torques decline and settle eventually to
their final equilibria, here reached after about 40 orbits.
This process of phase mixing inside the horseshoe region is called
torque saturation, and it occurs on timescales of the order of the libration time,
which is given by
\beq
\label{eq:tlib}
    \tau_{lib} = \frac{4 \, a_p}{3 \, x_s} \, {P_p}
\eeq
where $P_p$ is the orbital period of the planet, and $x_s$ the
half-width of the horseshoe region (see eq.~\ref{eq:horseshoe}).
In our case (for $q= 6 \times 10^{-5}$ and $H/r = 0.05$) the libration time 
is about $30 P$.
The different initial values of the unsaturated torques depend on the form of
the potential (eg. smoothing length), but note that the timescale to reach equilibrium
is similar in all cases.
This particular time behaviour of the torques and the process of saturation
has been described recently for isothermal discs by \citet{2009MNRAS.394.2283P}, see
also \citet{2006ApJ...652..730M}.

The two runs with the $\epsilon$-potential result in the most negative torque values, i.e.
the fastest inward migration (lower two curves in the upper panel).
While the total torques of the two $\epsilon$-potentials are nearly identical,
in the corresponding radial torque distribution
the cases are clearly separated, a fact which is due to cancellation effects
when adding the inner (positive) and outer (negative) contribution.
The slightly deeper cubic $r_{sm} = 0.8$ potential leads to a 
marginally decreased (in magnitude)
torque compared to the simulations with $\epsilon$-potential.
For the cubic $r_{sm} = 0.5$ potential we obtain an even less negative
equilibrium torque compared to all the other isothermal simulations.
As most of the corotation torque is generated in the vicinity of the planet, a change in the
density structure there (by deepening the potential) may have a significant impact on the
torque values.
We can compare our values of the torque with the well known formulae for the specific torque
in a 3D strictly {\it isothermal disc} as presented by \citet{2002ApJ...565.1257T}
\beq
\label{eq:tanaka}
     T^{tot}_0 = - \, f_\Gamma \, q \, 
           \left(\frac{H}{r}\right)^{-2} \, 
           \left(\frac{\Sigma a_p^2}{M_*}\right)  \,
          a_p^2 \Omega_p^2 
\eeq
with
\beq
        f_\Gamma = ( 1.364 + 0.541 \, \alpha_\Sigma )
\eeq
where $\alpha_\Sigma$ denotes the radial gradient of the surface density through 
$\Sigma \propto r^{- \alpha_\Sigma}$. For our standard parameter this formula gives
about $T^{tot}_0 = - 2.5 \, 10^{-5} a_p^2 \Omega_p^2$, which is, in absolute value,
about a factor $1.4 - 2.2$ times larger than our results.
We note however, that eq.~(\ref{eq:tanaka}) has been derived for constant temperature,
inviscid disc. 
The influence of viscosity on the torque has been studied by
\citet{2002A&A...387..605M}, who found that a reduction of the viscosity from our
used value of $10^{-5}$ to zero will fully saturate the 
(vortensity-related) corotation torque,
leading easily to a reduction of the total torque by a factor of two.
Additional simulations with much smaller viscosity 
(not shown here) 
indicate indeed that then the equilibrium torque is in good
agreement with eq.~(\ref{eq:tanaka}).

It seems at first surprising and unpleasing that the torques depend so much on the
treatment of the planetary potential. However, an
$\epsilon$-potential has an influence far beyond the Roche-radius of the
planet and certainly will change the torques acting on the planet. 
Here the corotation torques are affected most prominently and become
more and more positive as the smoothing length is lowered
\citep[see also][]{2009MNRAS.394.2283P}.
Nevertheless, in two-dimensional
simulations it has become customary to rely on $\epsilon$-potentials for the purpose
to take into account the finite thickness of the disc.
In a three-dimensional context, the more localised cubic-potential
with its finite region of influence may be more realistic. 
But for the isothermal case the increased potential depth leads to a 
very large accumulation of mass, as seen in Fig.~\ref{fig:Iso1dDensity}.
In such a case it will be very difficult to achieve convergence.
In the more realistic radiative case the situation is eased somewhat through
a temperature increase near the planet, as outlined below.

\begin{figure}
 \centering
 \includegraphics[width=0.9\linwx]{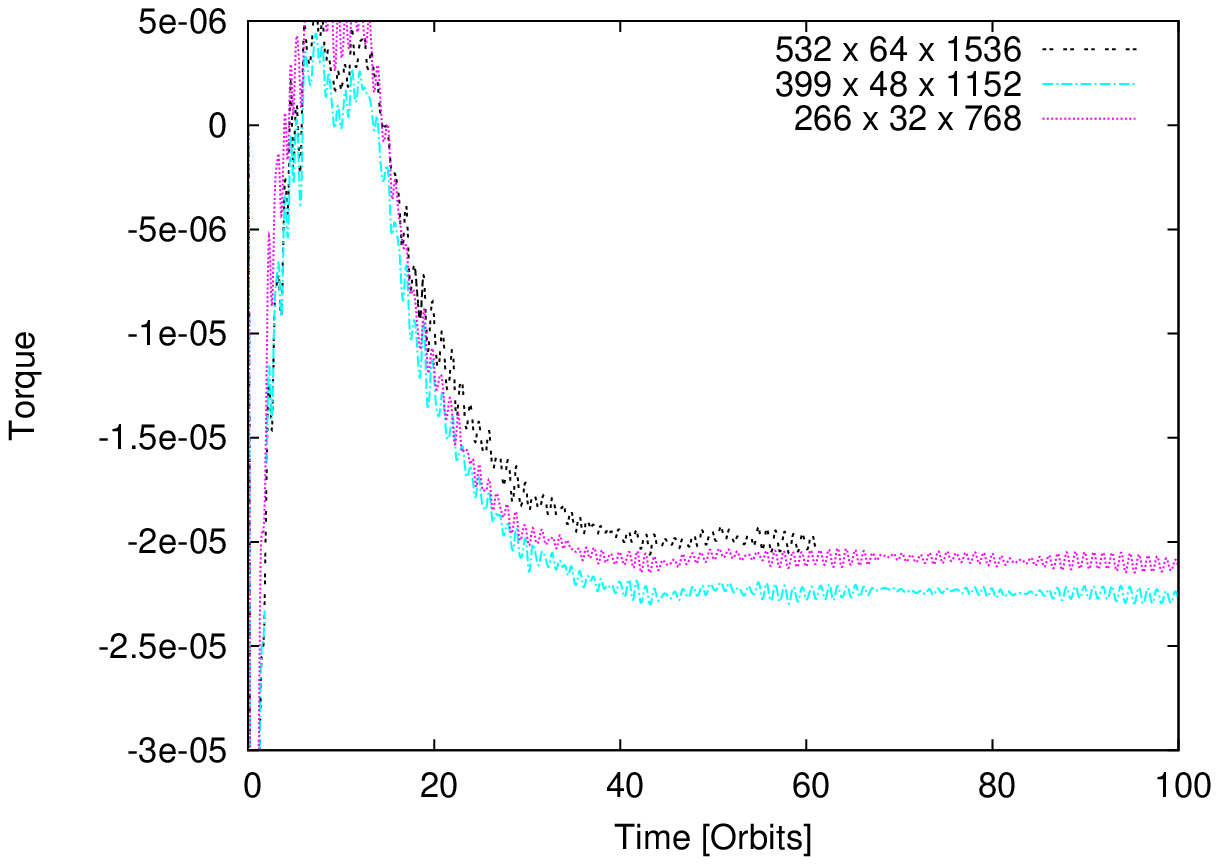} \\
 \includegraphics[width=0.9\linwx]{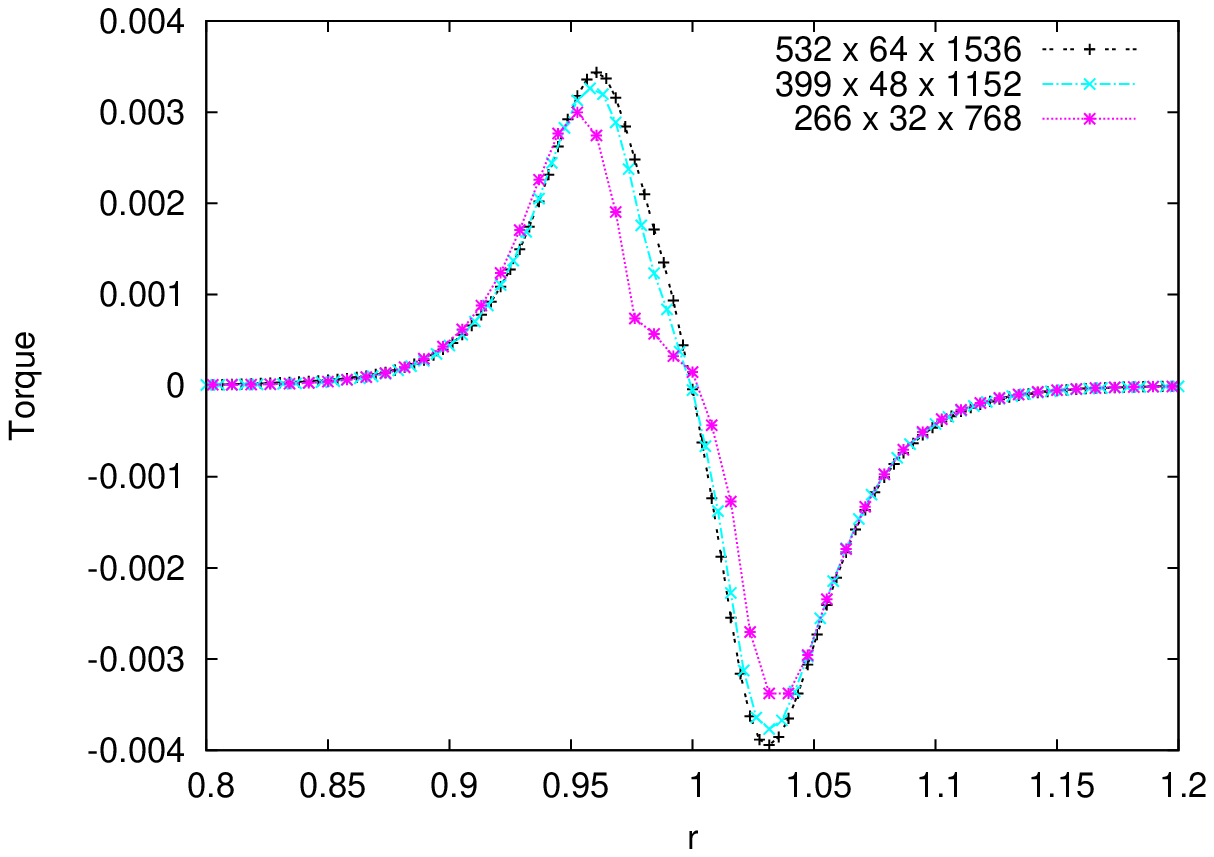}
 \caption{Specific torque acting on the planet using different grid resolutions for
  the isothermal case with $H/r = 0.037$.
  In all cases the cubic potential with $r_{sm} = 0.5$ has been used.
  {\bf Top}: Evolution of total torque with time.
  {\bf Bottom}: Radial variation of the specific torque density, at $t=80$ orbits
   for the standard and medium resolution, and at $60$ orbits for the high resolution.
   \label{fig:TorqueIsoBig}
   }
\end{figure}

To check numerical convergence we performed additional runs using a larger number of
gridcells. In Fig.~\ref{fig:TorqueIsoBig} we display the total torque versus time
and the radial torque density $\Gamma(r)$ for different grid resolutions, again
for the isothermal disc case.
In contrast to the previous plot we use here a slightly cooler disc with $H/r =0.037$,
as this matches more closely the results from the fully radiative calculations presented below.
In this case the initial unsaturated torques reach even positive values
due to the smaller thickness of the disc.
The grid resolution seems to be sufficient for resolving the
structures near the Roche-lobe. For the displayed $\Gamma(r)$ distribution in the lower panel
both cases are very similar and the higher resolution case is a bit smoother.

\subsection{Fully radiative discs}
\begin{figure}
 \centering
 \includegraphics[width=0.9\linwx]{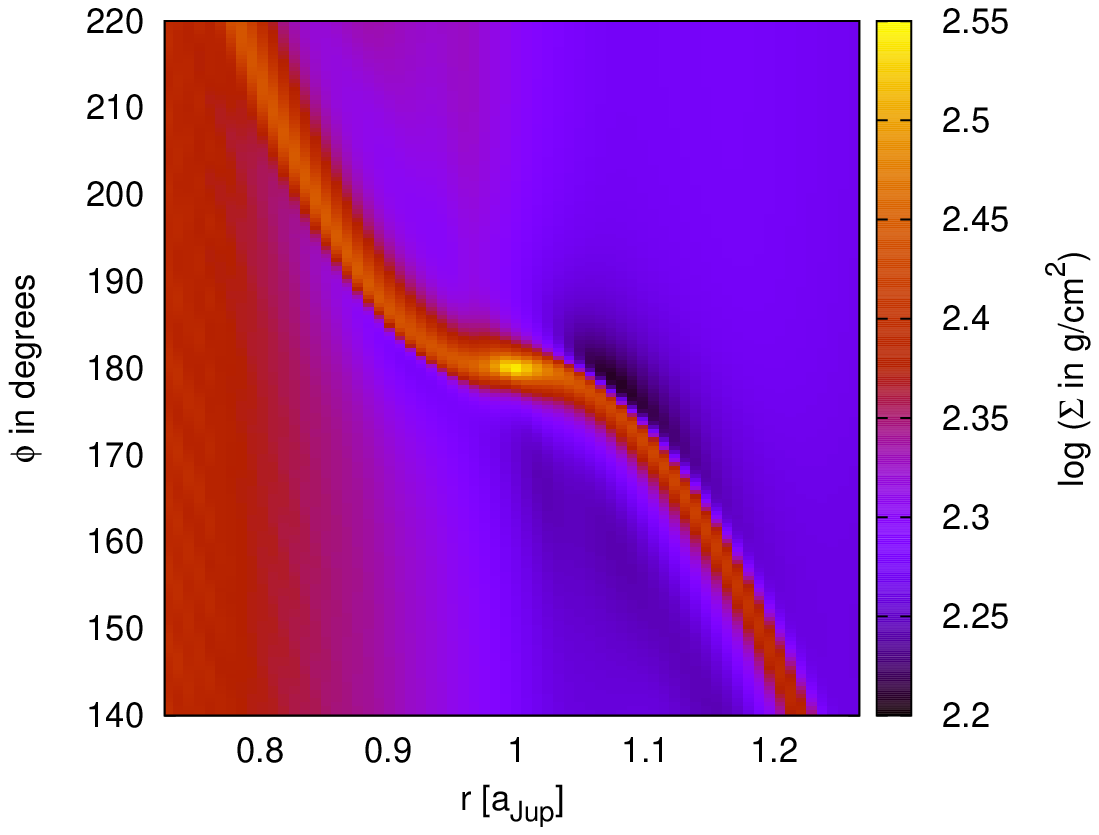}
 \includegraphics[width=0.9\linwx]{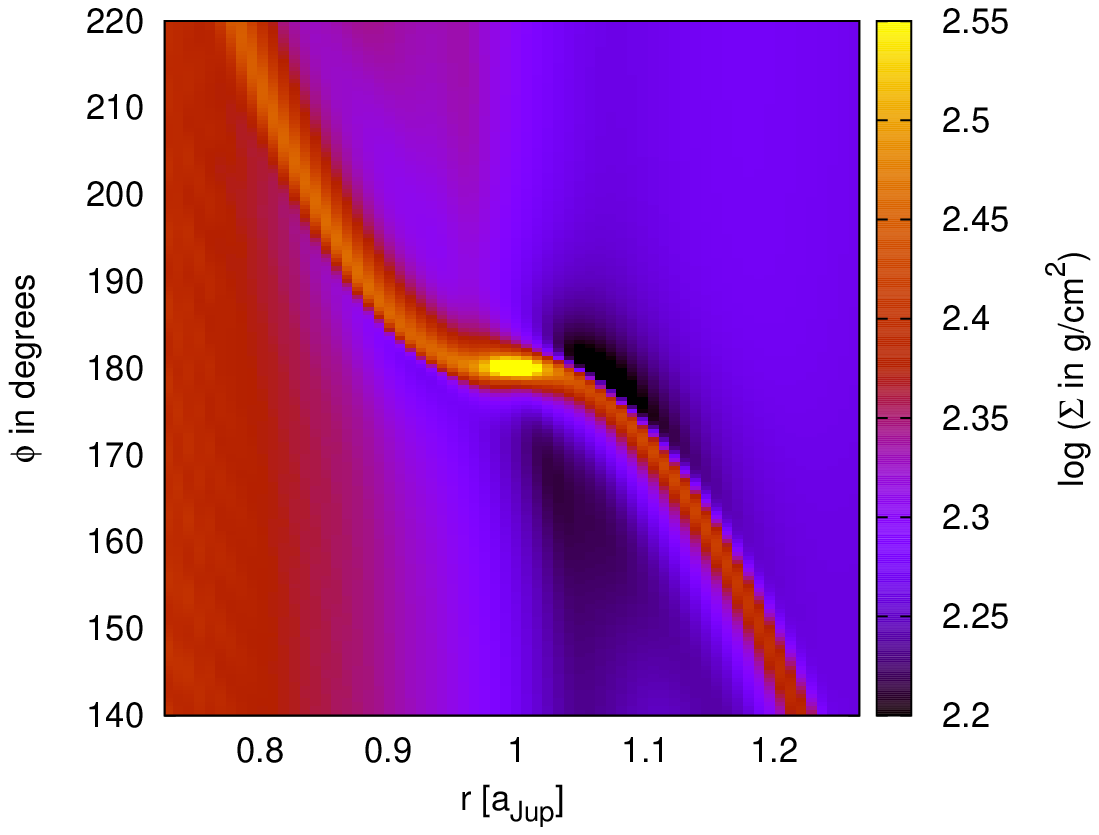}
 \includegraphics[width=0.9\linwx]{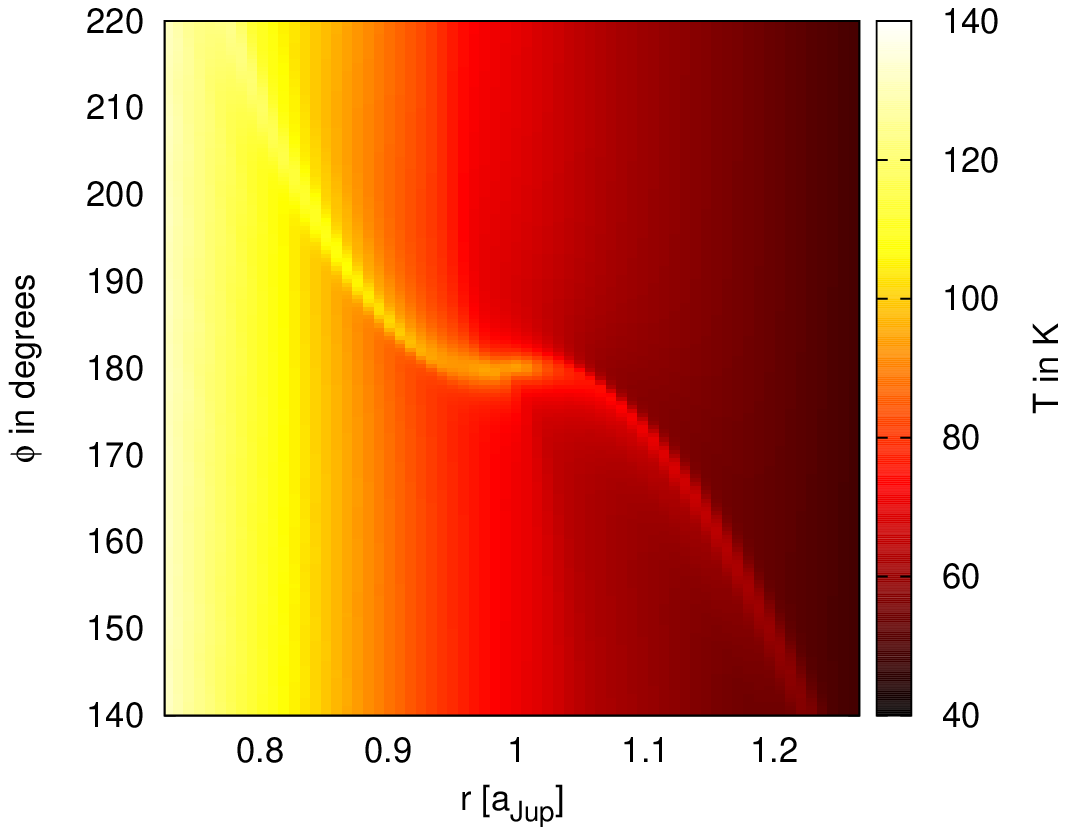}
 \includegraphics[width=0.9\linwx]{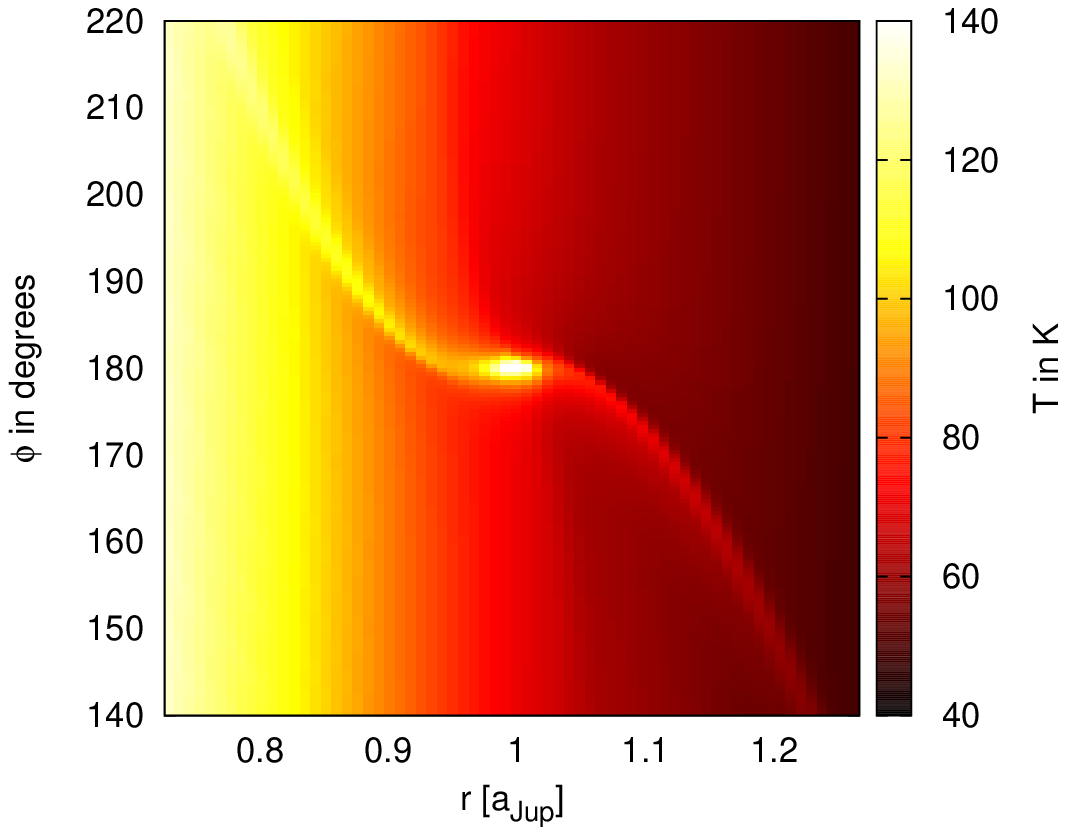}
 \caption{Surface density (upper two panels) and temperature in the equatorial plane (lower two) 
  for fully radiative simulations at 100 planetary orbits. 
 Displayed are results for the shallowest and deepest potential.
  Upper panels refer to the $\epsilon$-potential with $r_{sm}=0.8$, 
 and lower to the cubic-potential with $r_{sm}=0.5$, respectively.
   \label{fig:Full2d}
   }
\end{figure}

The simulations are started from the radiative disc in equilibrium as described above,
and are continued with an embedded planet of $20 M_{\rm earth}$.
The obtained equilibrium configuration for the surface density and midplane-temperature is
displayed in Fig.~\ref{fig:Full2d} after an evolutionary time of 100 orbits.
As in the isothermal case the density within the Roche lobe of the planet
is strongly enhanced for the deeper potentials, displayed are the two extreme
cases of our different potentials.
Comparing with the corresponding density maps of the
isothermal case \ref{fig:Iso2dDensity}, one can also observe slightly smaller opening angles of the spiral arms
in the radiative case. For identical
$H/r$ the sound speed would be $\sqrt{\gamma}$ times larger in the radiative case
leading to a larger opening angle.
Here, the effect is overcompensated by the reduced temperature (lower thickness)
in the radiative case. A different opening angle of the spiral arms will affect
the corresponding Lindblad torques acting on the planet. 

At the same time, a slight density enhancement is visible
'ahead' of the planet ($\varphi > 180^\circ$) at a slightly smaller radius ($r \lsim 1$).
This feature that is not visible in the isothermal case is caused
by including the thermodynamics of the disc.
Let us consider an {\it adiabatic situation} just after the planet has
been inserted into the disc, and follow material on its horseshoe
orbit (in the co-rotating frame) as it makes a turn from the outer disc
($r >1, \varphi > 180^\circ$) to the inner ($r <1$). 
The radial temperature
and density gradient imply for our ideal gas law a gradient in the entropy function $S$
in the disc through
\[
    S  \propto  \frac{p}{\rho^{\gamma}}.
\]
As shown above, in our simulations we find for the surface density
$\Sigma \propto r^{-1/2}$ as due to the assumption of a constant viscosity,
and the midplane temperature follows $T \propto r^{-1.7}$.
Due to this gradient in $S$ a parcel (coming from outside) has in our case a smaller
entropy than the inner disc which it is entering. Now the entropy
remains constant on its path, due to the adiabatic assumption.
Additionally, dynamical equilibrium requires that the pressure of the
parcel does not change significantly upon its turn, and entropy conservation 
then implies that the density has to increase.
At the same time the density `behind' the planet ($\varphi < 180^\circ$ and $r \gsim 1$)
will be lowered by similar reasoning.
Both components produce a positive contribution to this entropy-related corotation torque 
that acts on the planet, and which adds to the negative Lindblad torque and the positive 
vortensity-related corotation torque.
In truly adiabatic discs this effect will disappear after a few libration times
\citep{2008ApJ...672.1054B, 2008A&A...487L...9K} because the material, being within the horseshoe
region, will start interacting with itself, and the density and entropy
will be smeared out
due to the mixing, leading to the described torque saturation process.
Adding radiative diffusion will prevent this and keep the entropy-related torques unsaturated, and
a non-zero viscosity is also required.

In this fully radiative case the temperature within the Roche radius of the
planet has also increased substantially due to compressional heating of
the gas (lower two panels in Fig.~\ref{fig:Full2d}).
In addition, the temperature in the spiral arms is increased as well
due to shock heating.

\begin{figure}
 \centering
 \includegraphics[width=0.9\linwx]{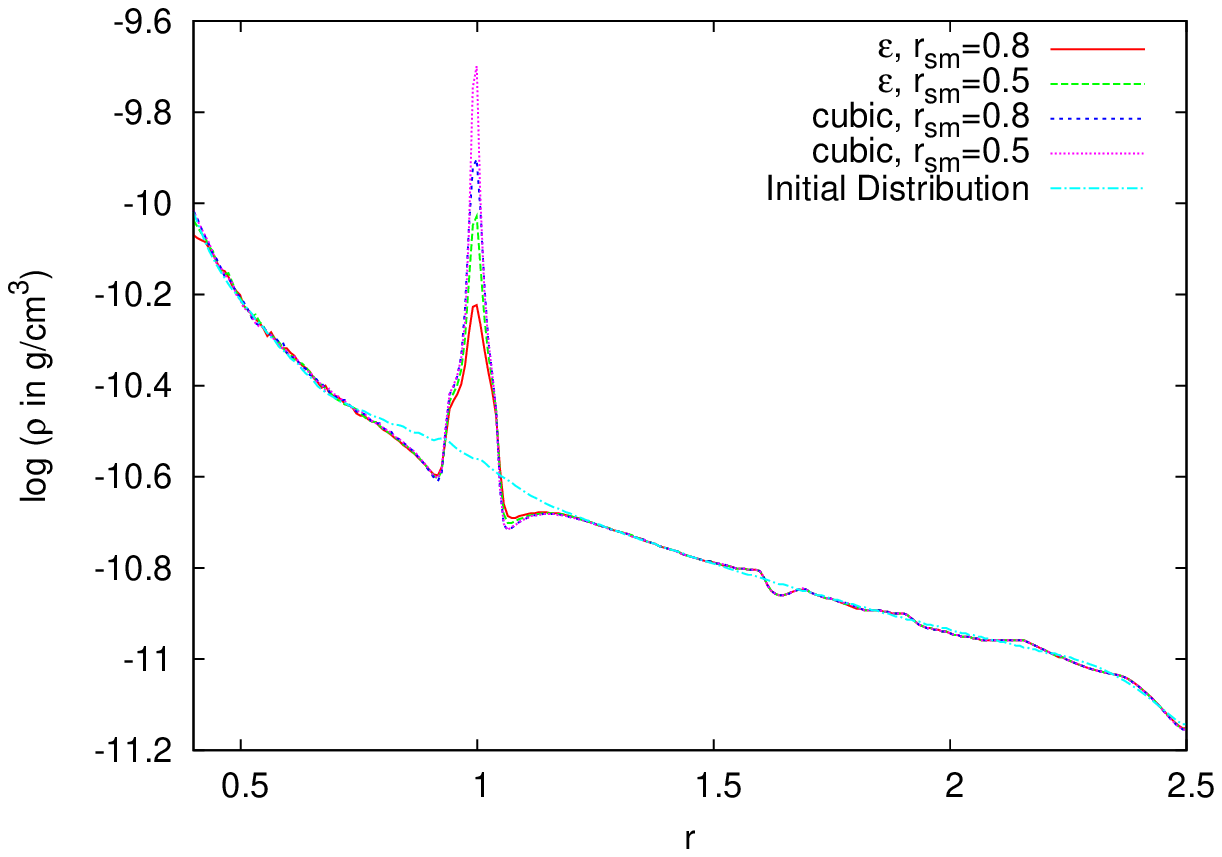}
 \includegraphics[width=0.9\linwx]{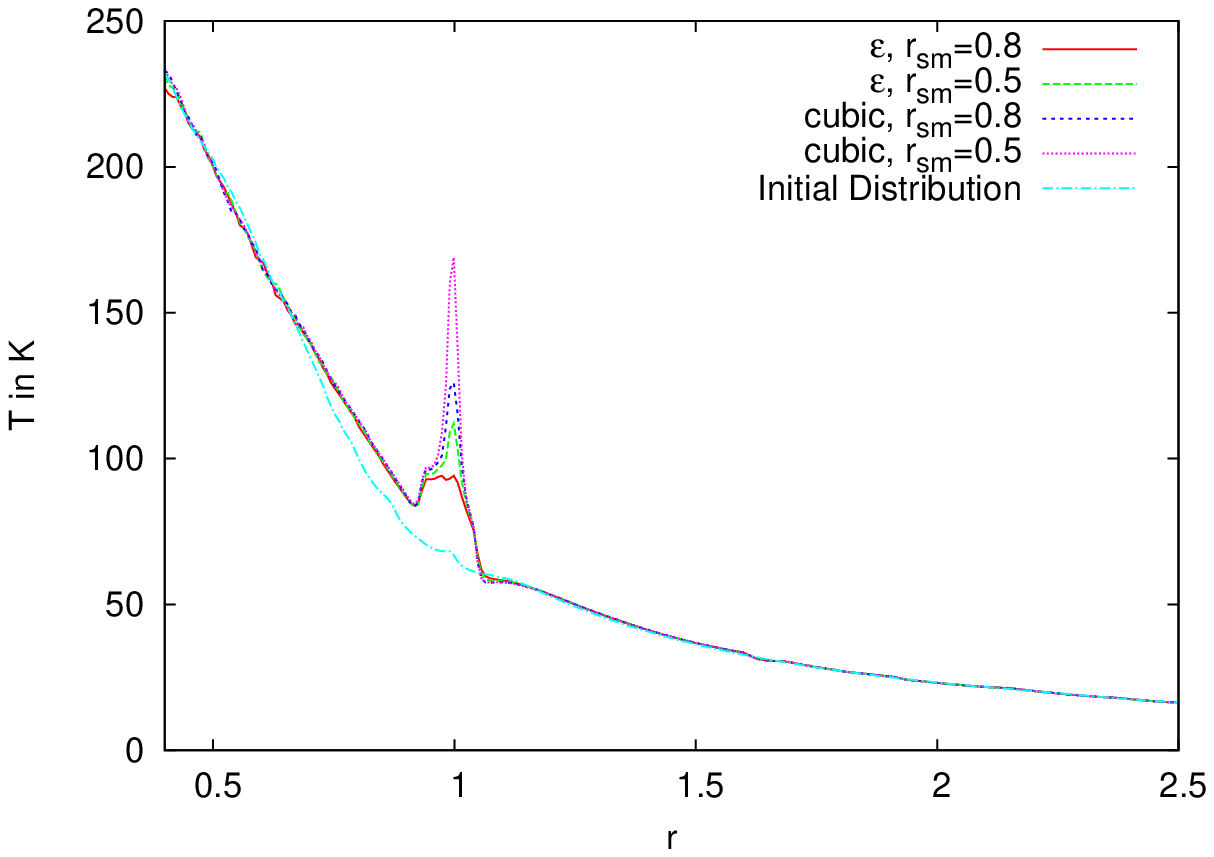}
 \caption{Radial density (top) and temperature (bottom) distribution in the 
  equator along a ray through the location of the planet for all 4 planetary potentials used,
   for the fully radiative case.
   \label{fig:Full1d}
   }
\end{figure}
The density and temperature runs in the disc midplane along a radial line at $\varphi=180^\circ$
cutting through the planet are displayed in Fig.~\ref{fig:Full1d}.
As in the isothermal case, deeper potentials lead to higher densities
within the Roche lobe. The increase is somewhat lower because now the temperature
is higher as well due to the compression of the material.
The larger pressure lowers the density in comparison to the isothermal case.
Interesting is that the maximum temperature 
is substantially higher than in the ambient disc even
for this very low mass planet of $20 M_{\rm earth}$.
Considering accretion onto the planet the increase in temperature 
might be even stronger due to the expected accretion luminosity.

\begin{figure}
 \centering
 \includegraphics[width=0.9\linwx]{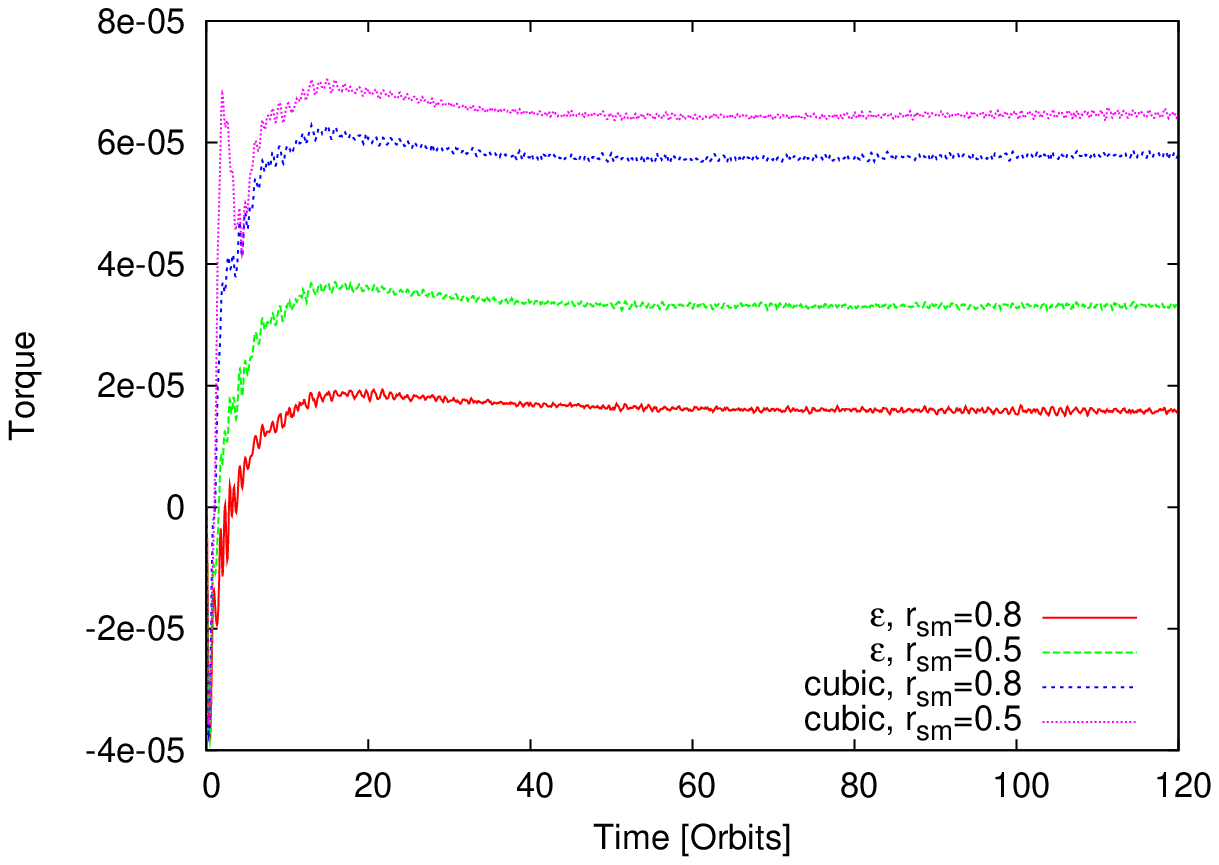}
 \includegraphics[width=0.9\linwx]{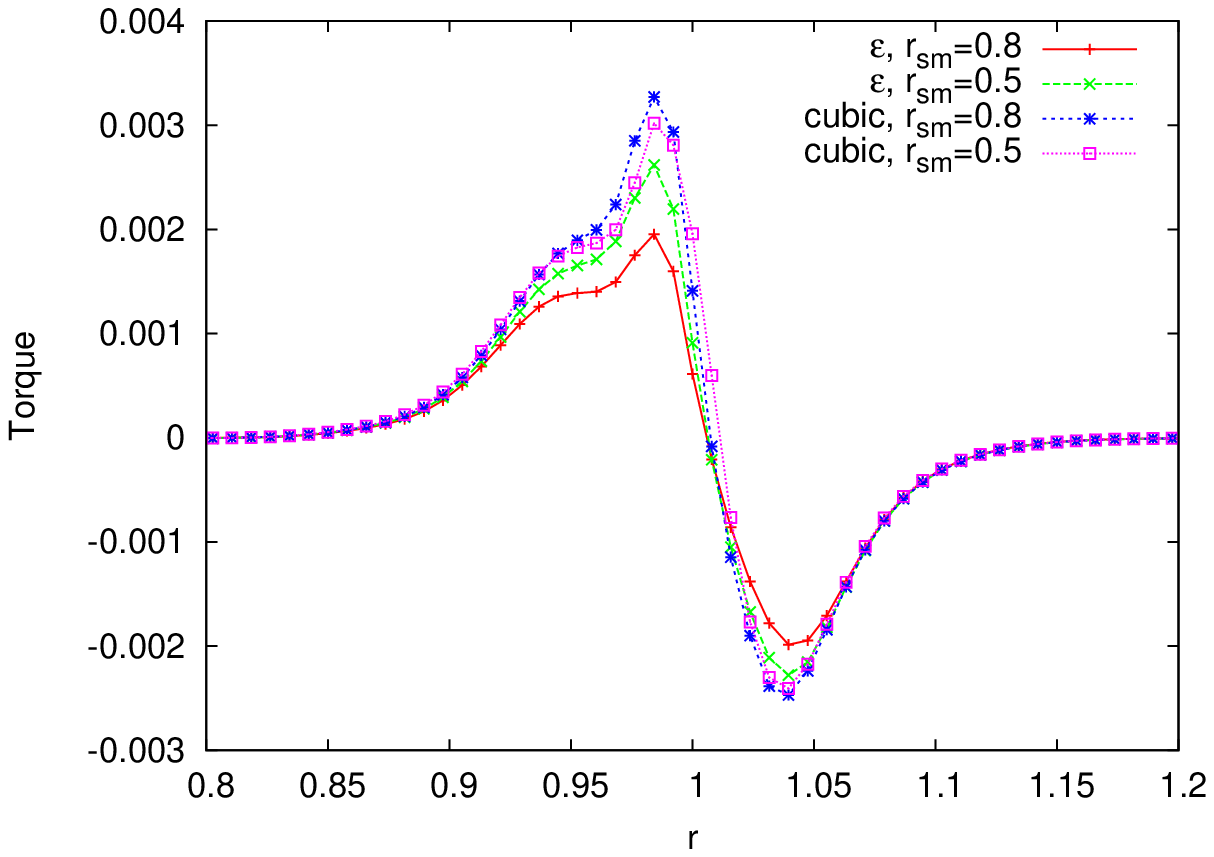}
 \caption{Specific torques acting on a 20 $M_{\rm earth}$ planet
  for different numerical potentials
  in the fully radiative case.
  {\bf Top}: Evolution of total torque with time.
  {\bf Bottom}: Radial variation of the specific torque for $t=80$ orbits.
   \label{fig:TorqueFull}
   }
\end{figure}

\begin{figure}
 \centering 
 \includegraphics[width=0.9\linwx]{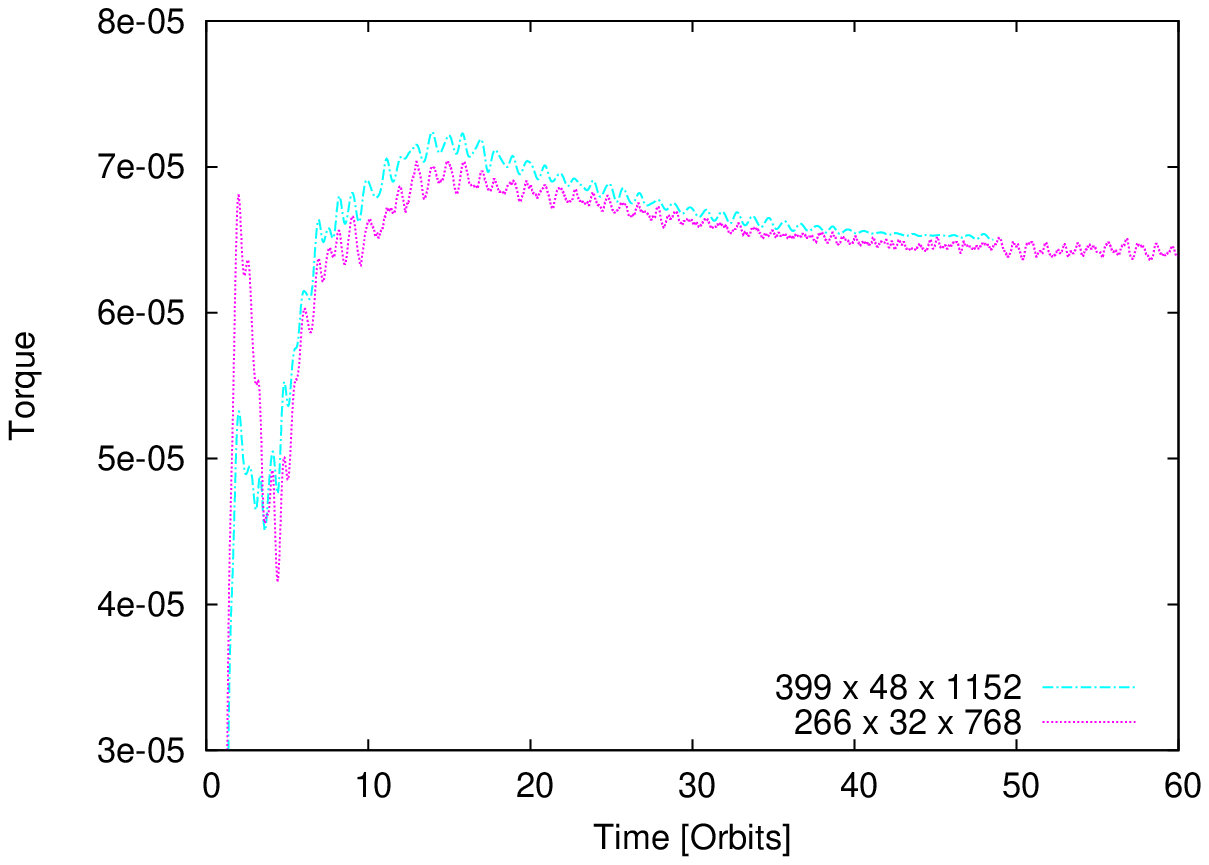} \\
 \includegraphics[width=0.9\linwx]{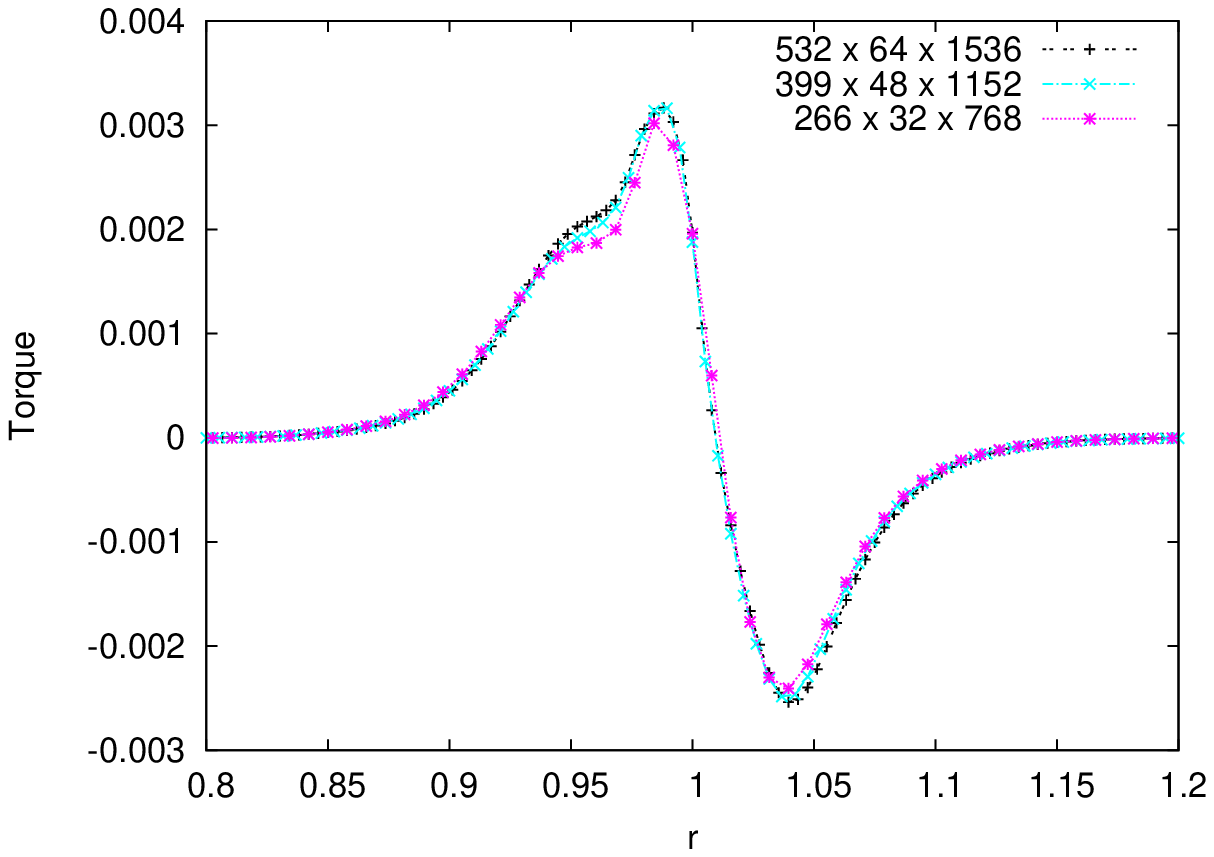}
 \caption{Specific torque acting on the planet using different grid resolutions for
  the fully radiative case.
  In all cases the cubic potential with $r_{sm} = 0.5$ has been used.
  {\bf Top}: Evolution of total torque with time.
  {\bf Bottom}: Radial variation of the specific torque density at $t=50$ orbits.
   \label{fig:TorqueFullBig}
   }
\end{figure}
\subsection{Torque analysis for the radiative case}
In the upper panel of Fig.~\ref{fig:TorqueFull} we display the time
evolution of the total specific
torque acting on a 20 $M_{\rm earth}$ planet for the full radiative case.
In contrast to the isothermal situation, all four potentials result now in a
positive total torque acting on the planet.
As in the previous isothermal runs, the torques
reach their maximum shortly after the onset of the simulations (between $t \approx 10-20$)
and then settle toward their final
value. In the corresponding isothermal case with $H/r =0.037$ the difference
between the initial positive unsaturated torque and the final saturated value has
been very pronounced (see Fig.~\ref{fig:TorqueIsoBig}). In contrast, in this fully radiative case the
inclusion of energy diffusion and the subsequent radiative cooling of the disc
will prevent saturation of the entropy-related corotation torque, resulting in a positive
equilibrium torque. Very similar results have been found previously in the fully radiative regime in
2D simulations \citep[][]{2008A&A...487L...9K}.
It is important to notice, that the two cubic-potentials (which are more
realistic in the 3D case) yield very similar
results. The more unrealistic $\epsilon$-potentials show rather strong deviations because,
due to their extended smoothing of the potential, they tend to weaken in particular the
corotation torques which originate in the close vicinity of the planet.
In the lower panel of  Fig.~\ref{fig:TorqueFull} the radial torque distribution
is displayed for the same 4 potentials. In comparison to the corresponding plot for
the isothermal $H/r =0.037$ (see Fig.~\ref{fig:TorqueIsoBig}) we notice
that the regular Lindblad part is slightly reduced in the
radiative case due to the larger sound speed.

\begin{figure}
 \centering 
 \includegraphics[width=0.9\linwx]{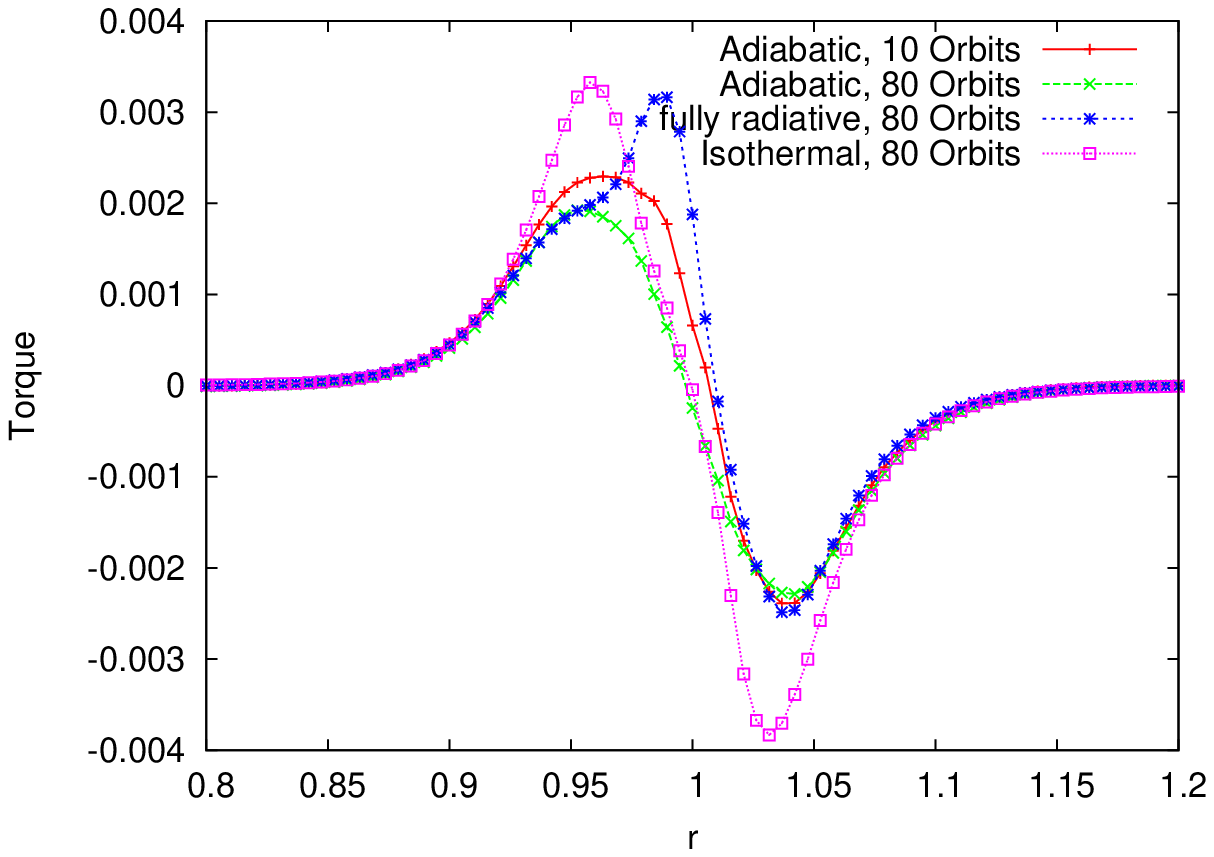} \\
 \caption{Radial variation of the specific torque acting on the planet
   using different thermodynamical disc models.
  In all cases the cubic potential with $r_{sm} = 0.5$ and standard resolution have been used.
  The adiabatic model at $t=10$ corresponds to the time where the corresponding total torque
  has its maximum. The models at $t=80$ have all reached their equilibria.
   \label{fig:TorqueVergleich}
   }
\end{figure}

Additionally, clearly seen is the additional positive contribution just inside $r=1$
which appears to be responsible for the torque reversal. This feature is caused by 
an asymmetric distribution of the density in the very vicinity of the planet,
see also Fig.~\ref{fig:Torque2DFullBig} below.
It pulls the planet gravitationally ahead, increasing its angular momentum, leading to
a positive torque. 
Above we argued that this effect may be due to the 
entropy-related corotation torque
of material moving on horseshoe orbits \citep[see also][]{2008ApJ...672.1054B}.
Due to the symmetry of the problem one might expect
a similar feature caused by the material moving from inside out.
However, there is no sign of this present in the lower panel of Fig.~\ref{fig:TorqueFull}.
To analyse this asymmetry, we performed additional simulations varying the grid
resolution and disc thermodynamics.
That the feature is not caused by lack of numerical resolution is demonstrated in
Fig.~\ref{fig:TorqueFullBig}, where results obtained with two different grids are displayed.
Both models show the same characteristic torque enhancement just inside the planet.
In Fig.~\ref{fig:TorqueVergleich} we compare the radial torque density of 
the isothermal and a new adiabatic model for $H/r=0.037$ with the fully 
radiative model, all for the cubic potential with $r_{sm}=0.5$ at intermediate resolution.
For the adiabatic case we show $\Gamma(r)$ at two different times. The first at $t=10$
when the torques are unsaturated, and the second at $t=80$ after saturation has occurred.
{Please note, that the isothermal and adiabatic models start from the same initial conditions
(locally isothermal), while the radiative model starts from the radiative equilibrium
without the planet.}
While the adiabatic model at $t=10$ shows signs of the enhanced torque just inside
the planet, there is no sign of a similar feature at a radius just outside of the planet.
Hence, this asymmetry of the entropy-related corotation torque is visible in both the adiabatic and
radiative case. Outside of the planet the adiabatic model  and the radiative
agree very well as $H/r$ is similar, while the isothermal model deviates due to the
different sound speed. 
Whether the location of the maximum in the torque density is identical in the radiative
and adiabatic case is hard to say from these simulations because in 3D adiabatic runs
the peak appears to be substantially broader with respect to corresponding 2D cases.
It has been argued by
\citet{2008ApJ...672.1054B} that it should occur exactly at the corotation radius,
which is shifted (very slightly) from the planet's location due to the pressure gradient in the disc. 
In our radiative simulations it seems that the maximum is slightly shifted inwards,
an effect which may be caused by adding radiative diffusion to the models and consider
discs in equilibrium. An issue that certainly needs further investigation.

\begin{figure}
 \centering
 \includegraphics[width=0.9\linwx]{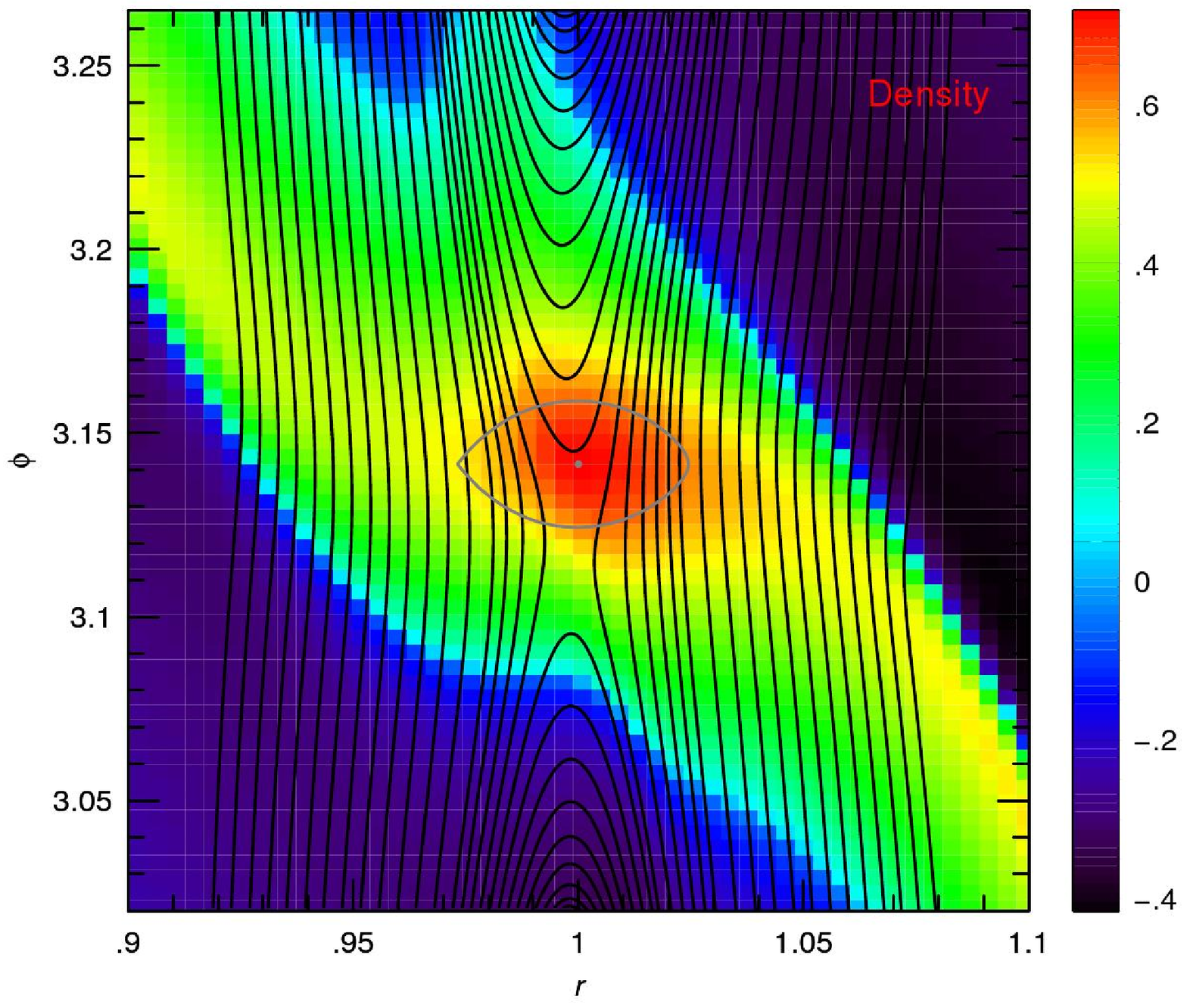} \\
 \includegraphics[width=0.9\linwx]{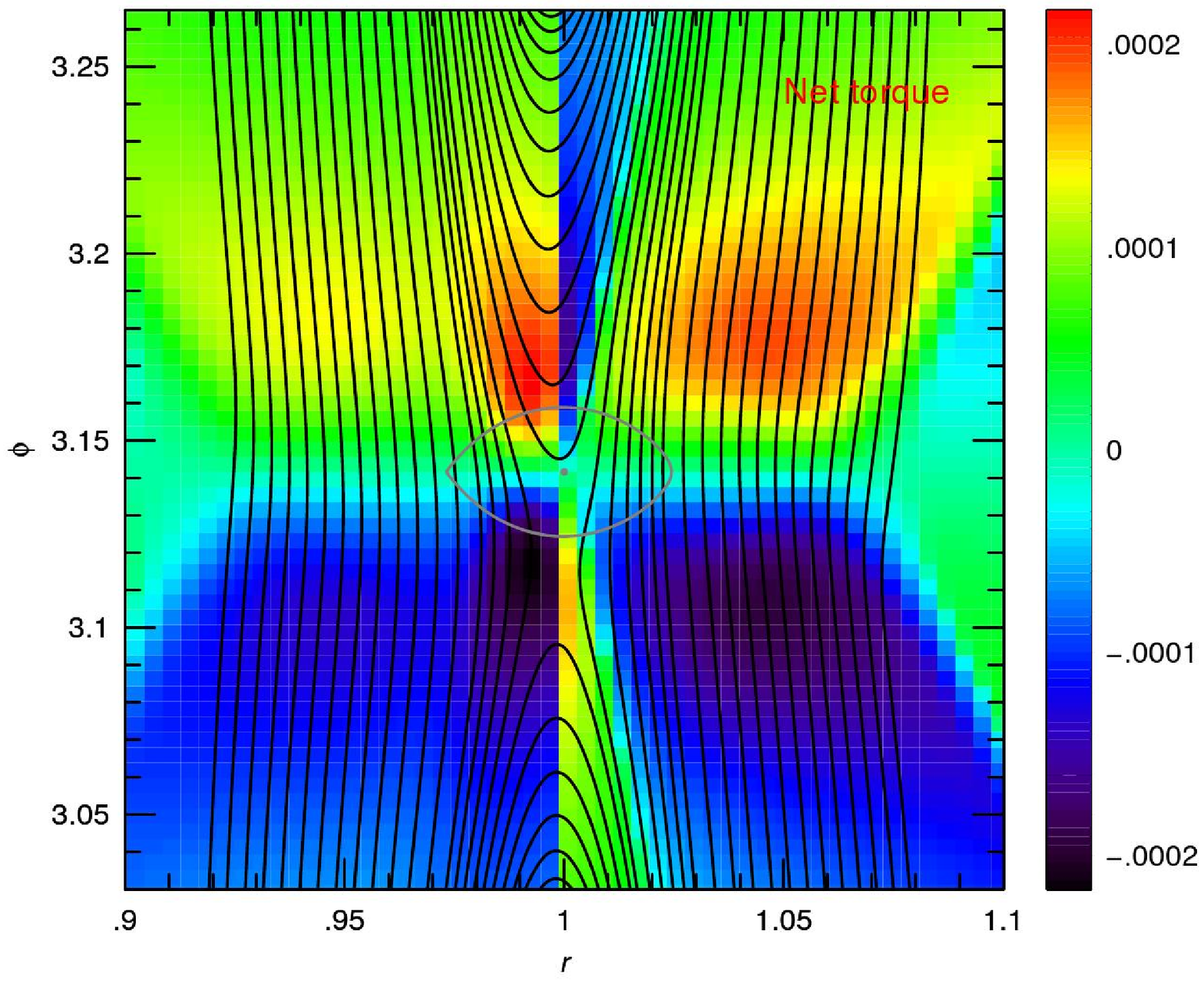}
 \caption{Results of a 2D fully radiative model using a resolution of $512 \times 1536$
  gridcells. The gray dot indicates the location of the planet, and the curve
   its Roche lobe. The solid black lines show the streamlines.
  {\bf Top}: Perturbed surface density with respect to the case without an embedded
   planet. The values are scaled as $\Sigma^{1/2}$.
  {\bf Bottom}: The net torque acting on the planet caused by the mass in each individual gridcell
    using the prescribed smoothed torque cutoff function with $b=0.8$, see explanation in text.
   The values are scaled as $(\tilde{\Gamma}^{\pm})^{1/2}$.
   \label{fig:Torque2DFullBig}
   }
\end{figure}
In Fig.~\ref{fig:Torque2DFullBig}  we present additional results of supporting 2D simulations
for the fully radiative case, as shown for lower resolution in \citet{2008A&A...487L...9K}.
All physical parameter are identical to our 3D fully radiative case.
The top panel shows the surface density distribution next to the planet. Seen is the
density enhancement just inside and ahead of the planet, and some indication
for a lowering outside and behind.
Please note, that the planet moves counter-clockwise into the positive $\phi$-direction.
i.e. upward in Fig.~\ref{fig:Torque2DFullBig}.
In the lower panel we display for each gridcell the net torque ($\tilde{\Gamma}^\pm$)
acting on the planet.
It is constructed by adding each cell's individual contribution to the torque
and that of the symmetric cell with respect to the planet location, i.e.
\beq
    \tilde{\Gamma}^\pm_{i,j}  = \pm \left[ \Gamma(r_i, \phi_{planet} + \phi_j) 
         + \Gamma(r_i, \phi_{planet} - \phi_j)
    \right].
\eeq
Hence, in absolute values the bottom half of the plot (with $\phi < \phi_{planet}$)
resembles exactly the top half ($\phi > \phi_{planet}$). 
The colours are chosen such that blue refers to negative $\tilde{\Gamma}^\pm_{i,j}$
and yellow/red to positive values.
The signs of $\tilde{\Gamma}^\pm_{i,j}$ are chosen such that the top left and lower right 
quadrant have the correct sign ($+$) and the other two are just reversed.
Due to this redundancy in the plot, only the upper left and lower right quadrant
should be taken into account to estimate global effects.
The mirroring process at the $\phi=\pi$-line 
(with the mirroring of the colour scale)
allows an easy evaluation and comparison of the individual contributions.
One can notice that the net torque will be positive due to excess material just ahead
and inside of the planet.
From the plot it is also clear that there exists indeed an asymmetry of the torques induced
by horseshoe material coming from outside-in versus material turning inside-out.
In the figure, there is only a weak indication of a marginal positive contribution
just below the planet.
In additional simulations for purely adiabatic discs with different (positive and negative)
entropy gradients which have either constant density or temperature it has  become
apparent that the asymmetry is caused by the entropy gradient. In the case of a negative
entropy gradient (as in our fully radiative model) the positive excess torque comes
from inside/ahead the planet, while for a positive gradient the negative excess torque
comes from outside/behind the planet. 
Whether the maximum of $\Gamma(r)$ lies at corotation \citep{2008ApJ...672.1054B} or is
slightly shifted when radiative effects are considered may deserve further studies.

\begin{figure}
 \centering
 \includegraphics[width=0.9\linwx]{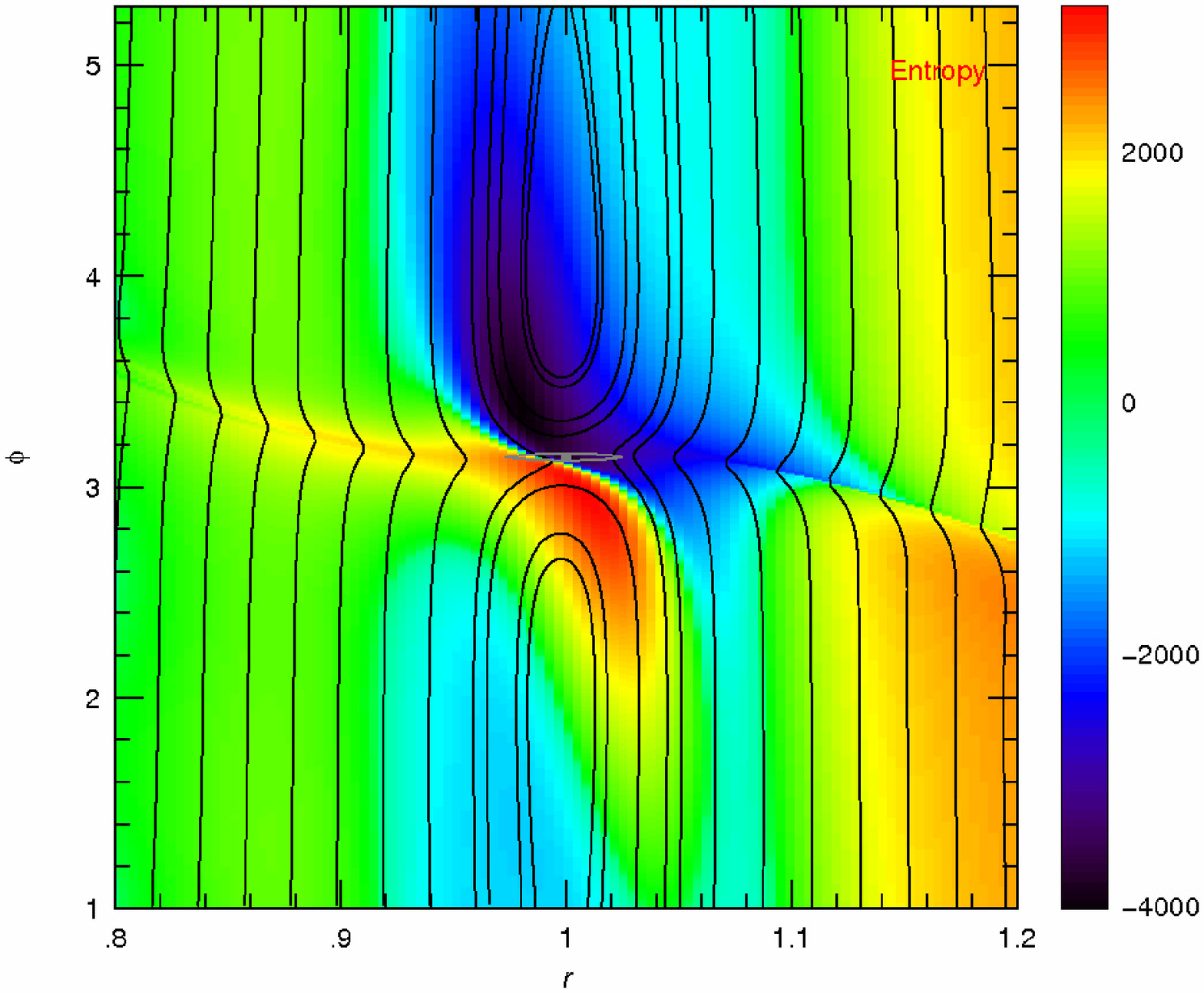} \\
 \includegraphics[width=0.9\linwx]{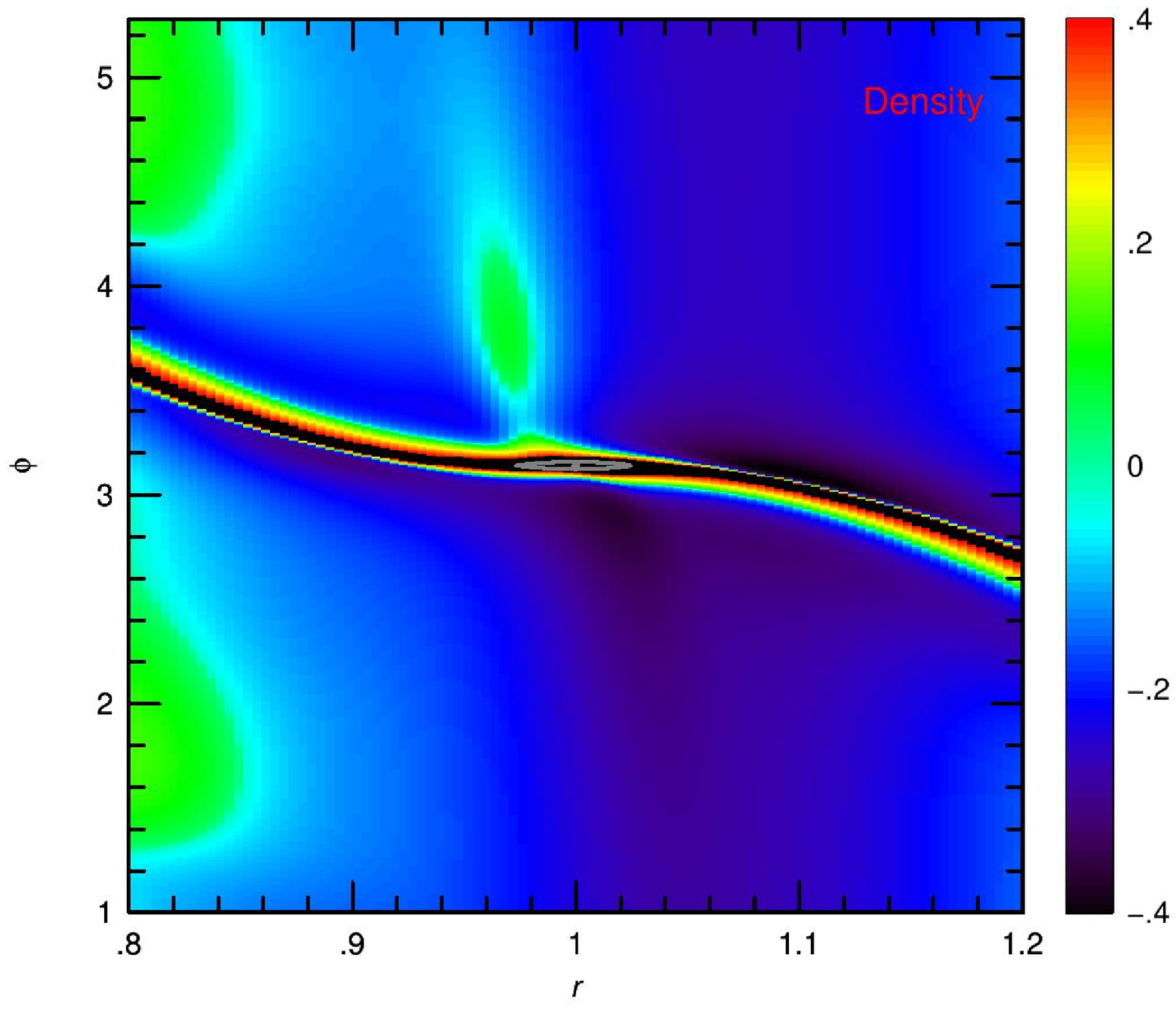} \\
 \caption{Perturbed entropy ({\bf Top}) and perturbed density ({\bf Bottom})
 for the 2D fully radiative equilibrium model
  using a resolution of $512 \times 1536$
  gridcells. The values are scaled as $\Sigma^{1/2}$ and  $S^{1/2}$, respectively.
   \label{fig:ent2dFullBig}
   }
\end{figure}
In Fig.~\ref{fig:ent2dFullBig} we show the perturbed entropy and density
in the 2D fully radiative model
in equilibrium, for a larger domain. 
Caused by the flow in the horseshoe region, there is an entropy minimum for larger
$\phi$ inside of the planet, and a maximum for lower $\phi$ outside, for $r+r_p$.
Both lie inside the horseshoe region and close to the separatrix.
The overall entropy distribution is very similar to that found by 
\citet{2008ApJ...672.1054B} for adiabatic discs shortly after the insertion of the planet.
Due to the included radiative diffusion this effect does not saturate in our case, and we
clearly support their findings even for the long term evolution. 
The disturbed entropy shows in fact a slight asymmetry 
(in amplitude) with respect to the planet, that reflects back on to to the density distribution
(bottom panel).

\subsection{Planets with different planetary masses}

Following the results obtained in the previous sections we adopt now
the cubic $r_{sm}=0.5$ planetary potential assuming that it is closest
to reality, and study the effects of planets with various masses
in fully radiative discs. Starting from the 2D radiative equilibrium state
(see section~\ref{subsec:initial}) we now place planets with masses ranging
from $5$ up to $100$ $M_{\rm earth}$ in the initially axisymmetric 3D disc.
The numerical parameters for these simulations are identical to those discussed
above.

\begin{figure}
 \centering
 \includegraphics[width=0.9\linwx]{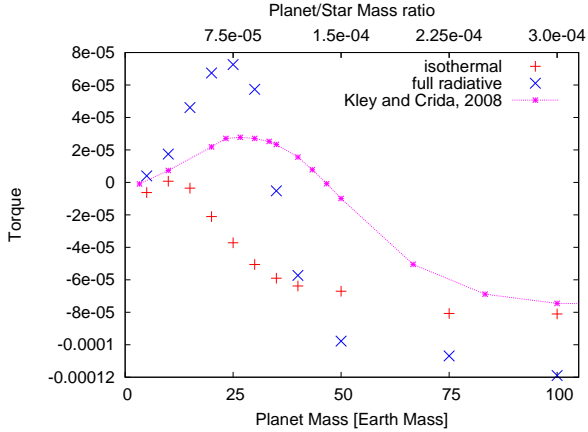}
 \caption{Specific torques acting on planets of different masses
 in the fully radiative (blue crosses) and isothermal (red plus signs) regime.
 Note, that the isothermal models are run for a fixed $H/r =0.037$.
  All torque values are displayed at a time when the equilibrium has been reached.
   \label{fig:TorqueMass}
   }
\end{figure}

In recent 2D simulations of radiative discs with embedded low mass planets
the torque acting on the planet depends on the planetary mass in such a
way that for planets with a size smaller than about $40$ earth masses the
total torque is positive implying outward migration
\citep[][]{2008A&A...487L...9K}. For larger
masses the forming gap reduces the contribution of the corotation torques,
and the results of the radiative simulations approach those of the fixed temperature
(locally isothermal) runs.
Our 3D simulations show indeed very similar results for planets in this
mass regime, see Fig.~\ref{fig:TorqueMass}. 
Planets in the isothermal regime
migrate inward with a torque proportional to the planet mass squared,
as predicted for low mass planets undergoing Type-I-Migration.
Note, that we use in these models the temperature distribution for a fixed $H/r=0.037$.
The values for the three lowest mass planets (with $5, 10, 15 M_{\rm earth}$)
are not as accurate due to the insufficient grid resolution, remember the 'kink'
in Fig.~\ref{fig:TorqueIsoBig} which refers to $20 M_{\rm earth}$ at standard resolution.
For the fully radiative disc the planets up to about $33 M_{\rm earth}$
experience a positive torque, while larger
mass planets migrate inward, due to the negative torque acting on them.

When comparing the 3D torques to the corresponding 2D values as obtained by
\citet{2008A&A...487L...9K} for the same disc mass and opacity law, we note two
differences:
{\it i)} The absolute magnitudes of the torques in the radiative case are
enhanced in the 3D simulations with respect to the corresponding 2D results,
resulting in even faster outward migration of the planets. 
This result can be explained by the reduced temperature (i.e. vertical thickness)
of the 3D disc with respect to the 2D counterpart (cf. Fig.~\ref{fig:radial}), as
a reduction in $H$ typically increases the torques \citep{2002ApJ...565.1257T}. 
{\it ii)} The turnover mass from positive to negative torques is reduced in the 3D
simulations. This effect is caused again by the reduced disc thickness, as
now the onset of gap formation ($R_H \approx H$) occurs for smaller planetary
masses.
The different form of the potential and the softening length 
may also play a role in explaining some of the differences observed 
between the 3D and 2D results.

\begin{figure}
 \centering
 \includegraphics[width=0.9\linwx]{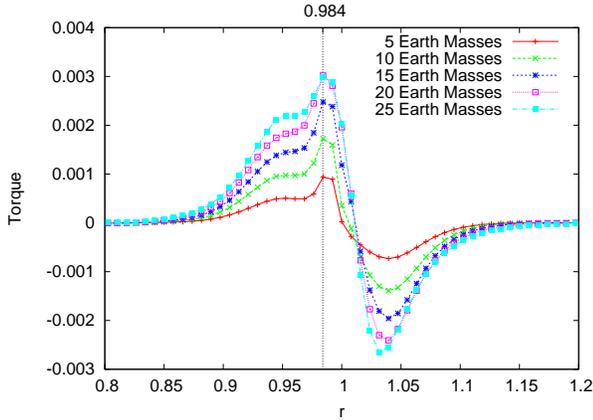}
 \caption{Radial torque distribution in equilibrium for various planet masses.
 The vertical dotted line indicates the location of the maximum. 
   \label{fig:RadTorqueMass}
   }
\end{figure}

Finally, in Fig.~\ref{fig:RadTorqueMass} we show that the position of the 
maximum of the radial torque density is independent of the planet mass, and is
therefore a result of the underlying disc physics.  
\section{Summary}
\label{sec:summary}
We have investigated the migration of planets in discs using
fully three-dimensional numerical simulations including radiative transport
using the code {\tt NIRVANA}.
For this purpose we have
presented and described our implementation of implicit radiative
transport in the flux-limited diffusion approximation, and secondly
our new {\tt FARGO}-implementation in full 3D.

Before embedding the planets we studied the evolution of axisymmetric,
radiative accretion discs in 2D. Starting with an isothermal disc model
having a fixed $H/r=0.05$, we find that for our physical disc parameter 
the inclusion of radiative transport
yields discs that are thinner ($H/r = 0.037$ at $r=1$).
We note that in the isothermal case the disc thickness is a chosen
input parameter, while in the fully radiative situation it 
depends on the local surface density and the chosen viscosity and opacity.
Interesting is here the direct comparison to the equivalent 2D models
using the same viscosity, opacity and disc mass \citep{2008A&A...487L...9K},
as displayed in Fig.~\ref{fig:radial}. Here, our new 3D disc yields
smaller temperatures (by a factor of 0.6-0.7) than the 2D runs.
Since the 2D simulations have to work with vertically averaged quantities,
it will be interesting whether it might be possible to adjust those as to yield
results in better agreement to our 3D results.

Concerning planetary migration we have confirmed the occurrence
of outward migration for planetary cores in radiative discs.
As noticed in previous research, the effect is driven by a
radial entropy gradient across the horseshoe region in the disc,
that is maintained by radiative diffusion.
Our results show that planets below the turnover mass of about
$m \approx 33 M_{\rm earth}$ migrate outward while larger masses drift inward.
The reduced temperature in the 3D versus 2D runs has direct influence on the magnitude
of the resulting torques acting on the planet.
As the disc is thinner in 3D the resulting torques, corotation as well
as Lindblad, are also stronger. The turnover mass from outward to inward
migration is slightly reduced as well for the 3D disc since the smaller
vertical thickness allows for gap opening at lower planet masses.
Due to the reduced temperature in the 3D case,
the spiral waves have a slightly smaller opening
angle compared to the isothermal case.

Another interesting, partly numerical, issue that we have addressed
concerns the influence that the smoothing of the planetary potential has
on the density structure in the vicinity of the planet and the speed
of migration.
We have compared the standard $\epsilon$-potentials in contrast with so-called
cubic-potentials. The results indicate,  that a deeper (cubic) potential
results in a higher density inside the planet's Roche
lobe for isothermal, as well as radiative discs. Since the potential
depth influences the density in the immediate vicinity of the planet
the resulting torques show some dependence on the chosen smoothing length.
We note that for the more realistic cubic-potential, changing  the smoothing
from $0.8$ to $0.5$ does not alter the results for the fully radiative
simulations considerably, and runs at different numerical resolutions have
indicated numerical convergence. Hence, we believe that this value is suitable
for performing planet disc simulations in 3D. The usage of an $\epsilon$-potential
cannot be recommended in 3D.
Outside of the planet's Roche lobe one can hardly notice a difference in the
density structure.
Since the cubic-potential agrees with the true planetary potential
outside of the smoothing-length $r_{sm}$, it is of course desirable to choose
this transition radius as small as possible, but the achievable 
numerical resolution always will set a lower limit. We found $r_{sm}=0.5$
a suitable value for our grid resolution and used that in our
parameter studies for different planet masses.
We note, that independent of the chosen form of the potential the 
outward migration of planetary cores seems to be a robust result for
radiative discs. As the magnitude (and direction) of the effect depends on the viscosity
and opacity, further studies, investigating different radial locations in the
disc, will be very interesting.

The outward migration of  planet embryos with several earth masses certainly
represents a solution to the too rapid inward migration found in this mass
regime of classical type-I migration. Growing planets can spend 
more time in the outer disc regions and move then later via type-II migration
towards the star.
On the other hand it may be difficult to reconcile this finding with
the presence of the discovered Neptune-mass planets that reside closer
to the central star ($a\approx 0.1$AU), but still too far away to be ablated by stellar
irradiation.

As seen in our simulations, parts of the disc can display convection due
to the form of the opacity law used. It will be certainly interesting in the future
to analyse what influence the convective motions have on the migration properties of the
embedded protoplanets. Additionally, fully radiative 3D-MHD simulations
are definitely required to judge the efficiency of this process in
turbulent discs.

\begin{acknowledgements}

Very fruitful discussions with Aur\'elien Crida and Fred\'eric Masset
are gratefully acknowledged. H. Klahr and W. Kley acknowledge the
support through the German Research Foundation (DFG) through
grant KL 650/11 within the Collaborative Research Group FOR 759:
{\it The formation of Planets: The Critical First Growth Phase}.
B. Bitsch has been sponsored through the German D-grid initiative.
The calculations were performed on systems of the Computer centre of the
University of T\"ubingen (ZDV) and systems  operated by the ZDV on behalf of bwGRiD,
the grid of the Baden  W\"urttemberg state.
Finally, we gratefully acknowledge the very helpful and constructive comments of an
anonymous referee that inspired us to perform more detailed analysis.

\end{acknowledgements}

\appendix
\section{Fargo algorithm}
\label{app:fargo}
Multi-dimensional simulations of accretion discs that include the $\varphi$-direction
typically suffer from severe timestep limitations. This is due to the
fact that the azimuthal velocity $u_\varphi$ is Keplerian and falls off with
radius. Hence, the innermost rings determine the maximum timestep allowed even
though the region of interest lies much further out.
One suggestion to resolve this issue is given by the
{\tt FARGO} algorithm which stands for ``Fast Advection in Rotating Gaseous Objects''
\citep{2000A&AS..141..165M}. It has originally been developed for 2D disc simulations
in a cylindrical coordinate system, for details of the implementation see
\citet{2000A&AS..141..165M,2000ASPC..219...75M}.
Here, we briefly describe our extension to three spatial dimensions in spherical
polar coordinates.

The basic method relies on a directional splitting of the advection part,
where first the radial and meridional (in $\theta$ direction) advection are 
performed in the standard way.
To calculate the azimuthal part we follow \citet{2000A&AS..141..165M} and split
the angular velocity into three parts:
From the angular velocity of each grid cell $\omega_{i,j,k} = (u_{\varphi})_{i,j,k}/r_i$ 
first an average angular velocity $\bar{\omega}_i$ is calculated for each radial ring $i$,
which is obtained here by averaging over the azimuthal (index $k$) and vertical 
(index $j$) direction
\beq
\label{eq:average}
   \bar{\omega}_i  = \frac{1}{N_\varphi \, N_\theta} \, \sum_{j,k} \omega_{i,j,k}
\eeq
where the summation runs of all azimuthal and meridional gridcells, and
$N_\varphi, N_\theta$ denote the number of these gridcells, respectively.
We note, that the summation over the vertical direction with index $j$ is
not required at this point. In our case, for a thin disc where the angular
velocity does not vary much with height, the vertical
averaging simplifies matters somewhat.
From this, one calculates an integer-valued shift quantity
\beq
    n_i = {\tt Nint} \left( \bar{\omega}_i \Delta t/\Delta \varphi \right),
\eeq
where {\tt Nint} denotes the nearest integer function.
This corresponds to a transport by the angular 'shift velocity'
\beq
      \omega_i^{SH} = n_i \frac{\Delta \varphi}{\Delta t}.
\eeq
Then we calculate the constant residual velocity of each ring
\beq
   \omega_i^{cr}= \bar{\omega}_i - \omega_i^{SH},
\eeq
and finally the residual velocity for each individual  gridcell
\beq
   \omega_{i,j,k}^{res} = \omega_{i,j,k}  - \bar{\omega}_i.
\eeq
Rewritten, we find for the angular velocity the following expression
\beq
   \omega_{i,j,k} \, = \,  \omega_{i,j,k}^{res}  + \omega_i^{cr}  + \omega_i^{SH} 
\eeq
The advection algorithm in the $\varphi$-direction proceeds now in three
steps. In the first two steps all quantities are advected using the standard
advection routine with the transport velocities $\omega_i^{cr}$ and $\omega_{i,j,k}^{res}$ and
then all quantities are shifted by the integer values $n_i$ in each ring $i$
which corresponds to a transport velocity $\omega_i^{SH}$. Using this splitting, the
transport velocities in the advection part are given by the 
two residual velocities $\omega_i^{cr}$ and $\omega_{i,j,k}^{res}$,
which are typically much smaller than $\omega_{i,j,k}$.
Hence, the time step limitation for the azimuthal direction is determined by the local
variation from the mean azimuthal flow in the disc which is typically much smaller than
the Keplerian value. 
In our case of a 3D disc the time step criterion is first given by the
normal CFL-criterion as presented for example in \citet{1992ApJS...80..753S}
where the angular velocity $\omega_{i,j,k}$ is just replaced by the
residual cell values $\omega_{i,j,k}^{res}$ and $\omega_i^{cr}$. This change provides the
major reduction in the transport velocity and a substantial increase in
the time step size.
An additional time step limitation is given by the requirement that the shift
should not disconnect two neighbouring grid cells in the radial and in the meridional
direction \citep{2000A&AS..141..165M}.
Here, this additional limit on the time step reads
\beq
     \Delta t_{shear} = 0.5 \, \min_{i,j,k} \left\{ 
       \frac{\Delta \varphi}{\left|\omega_{i,j,k} - \omega_{i-1,j,k}\right|}, \, 
       \frac{\Delta \varphi}{\left|\omega_{i,j,k} - \omega_{i,j-1,k}\right|}
       \right\} 
\eeq
The second restriction is only necessary in the case, where the above vertical averaging
in Eq.~\ref{eq:average} has not been performed.
The sequencing of the advection sweeps in the {\tt FARGO} algorithm has to be such
that the azimuthal sweep comes at the end, hence in our simulations
we use always: radial, meridional, and finally azimuthal.

In a staggered mesh code such as our {\tt NIRVANA} code, that is essentially
based on the {\tt ZEUS} method, an additional complication arises
in the straightforward application of the {\tt FARGO} method,
due to the fact that the velocity variables are located at the
cell interfaces and not at the centres.
Hence, the corresponding 'momentum cells' for the radial and meridional momenta
($\rho u_r$ and $\rho r u_\theta$) are shifted with respect to the standard density
cells by half a gridcell in the radial or meridional direction, respectively.
To apply the {\tt FARGO} method one has first to split all the momentum cells
in two halves, use the algorithm outlined above on each of the halves,
and then combine them again afterwards to calculate from the
updated momenta the new velocities on the interfaces.
This leads of course to an overhead in the simulation cost which is counterbalanced
however by the much larger time step.

\section{Radiative transport}
\label{app:fld}
Here, we outline briefly the method to solve the flux-limited diffusion
equation in 3D. Radiative transport is treated as a sub-step of the integration
procedure. In equilibrium viscous heating is balancing radiative diffusion
and to ensure this also numerically we incorporate the dissipation into
this sub-step.
Using the appropriate parts of the energy equation (\ref{eq:energy}) and
the flux (\ref{eq:raddif}) we obtain a diffusion equation for the gas temperature.
\beq
 \pdoverdt{T}  =
   \frac{1}{c\subscr{v} \rho} \, \left[
  \nabla \cdot  \, D  \nabla T  +  Q^+  \right]
 \label{eq:Tdif}
\eeq
where the diffusion coefficient is given by
\beq
 D =  \frac{\lambda c 4 a_R T^3}{\rho (\kappa + \sigma)},
 \label{eq:Diff}
\eeq
and $Q^+$ denotes the viscous dissipation that is added to the system.
The flux-limiter $\lambda$ depends on the local physical state of the gas and
approaches $\lambda = 1/3$ in the optically thick parts and reduces the
flux to $F = c a_R T^4$ in the optically thin parts. Here we use
an expressions for $\lambda$ as given in \citet{1989A&A...208...98K}.

A straight forward finite difference form of Equation (\ref{eq:Tdif})
in Cartesian Coordinates is given by
\bea
\label{eq:Tdiscret}
\frac{T^{n+1}_{i,j,k} - T^n_{i,j,k}}{\Delta t}
  & = &
\frac{1}{(c\subscr{v} \rho)_{i,j,k}}
\left[  \right.  \\
   \, & \, & 
\frac{1}{\Delta x}    \left(
\bar{D}^x_{i+1,j,k} \frac{T_{i+1,j,k} - T_{i,j,k}}{\Delta x}
      - \bar{D}^x_{i,j,k} \frac{T_{i,j,k} - T_{i-1,j,k}}{\Delta x}
\right)  \nonumber \\
   &  + & 
\frac{1}{\Delta y}    \left(
\bar{D}^y_{i,j+1,k} \frac{T_{i,j+1,k} - T_{i,j,k}}{\Delta y}
      - \bar{D}^y_{i,j,k} \frac{T_{i,j,k} - T_{i,j-1,k}}{\Delta y}
\right) \nonumber \\
   & + & 
\frac{1}{\Delta z}   \left(
\bar{D}^z_{i,j,k+1} \frac{T_{i,j,k+1} - T_{i,j,k}}{\Delta z}
      - \bar{D}^z_{i,j,k} \frac{T_{i,j,k} - T_{i,j,k-1}}{\Delta z}
\right)
\left.
\right]   \nonumber
\eea
In orthogonal curvilinear coordinates additional geometry
terms have to be added in the above equation.
Here $\bar{D}^x_{i,j,k}$ denotes
\beq
   \bar{D}^x_{i,j,k}  =  \frac{1}{2} \left( D_{i,j,k} + D_{i-1,j,k} \right)
\eeq
and so forth. 

\begin{figure}[ht]
\centerline{
\resizebox{0.75\linwx}{!}{%
\includegraphics{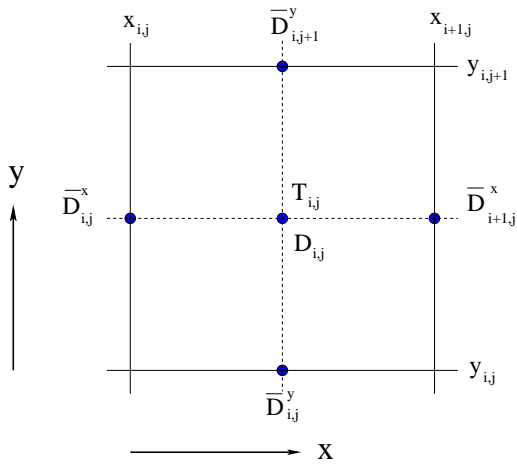}}
}  
  \caption{
The grid structure used in a staggered grid code for a two-dimensional example.
Shown is the cell $i,j$ where the coordinates of the cell centre [$x^c_{i}, y^c_j$]
is given by [$1/2(x_{i,j}+x_{i+1,j}), 1/2(y_{i,j}+y_{i,j+1})$].
The temperature $T_{i,j}$ is located at the cell centre where also the diffusion
coefficient $D$ is defined.
The averaged diffusion coefficients $\bar{D}$ are defined at the cell interfaces.
  }
   \label{fig:tempgrid}
\end{figure}

The grid structure from which Eq.~(\ref{eq:Tdiscret}) follows is outlined in the two-dimensional
case in Fig.\ref{fig:tempgrid}. One has to
keep in mind that the temperature (being a scalar) is defined in the cell centre, while
the values of $\bar{D}^x_{i,j,k}$ are defined at the cell interfaces.

In Eq.~(\ref{eq:Tdiscret}) no time levels are specified on the right hand side.
For explicit differencing the time level should be $t^{n}$ such that
the new temperature on the left $T^{n+1}$ is entirely given by the old values
$T^{n}$ at time $t^{n}$.
This might lead to very small timesteps since the timestep limitation
is approximately given by
\beq
   \Delta t  \leq  \min_{i,j,k}
  \left( \frac{\Delta x^2, \Delta y^2, \Delta z^2}{ \tilde D_{i,j,k}} \right)
\eeq
where $\tilde D$ is given by $D/(\rho c_v)$.

Hence, often an {\it implicit} version of the equation has to be used,
where all the temperature values $T_{i,j,k}$  on the r.h.s. are evaluated at
the new time $t^{n+1}$ or an arithmetic mean between new and old times.
Even though the diffusion coefficients may depend on temperature, we always
evaluate those at the old time $t^n$. Otherwise this would lead to a non-linear matrix
equation.

Collection all the terms in eq.~(\ref{eq:Tdiscret}) this leads to a linear system of equations
with the form
\bea
   A^x_{i,j,k} T_{i-1,j,k}  +  C^x_{i,j,k} T_{i+1,j,k}   +
   A^y_{i,j,k} T_{i,j-1,k} &  + &  C^y_{i,j,k} T_{i,j+1,k}   \nonumber  \\
   A^z_{i,j,k} T_{i,j,k-1}  +  C^z_{i,j,k} T_{i,j,k+1}   +
   B_{i,j,k} T_{i,j,k}  & = &
   R_{i,j,k}
 \label{eq:tmat}
\eea
where the superscript $n+1$ has been omitted on the left hand side.
The coefficients $A^x_{i,j,k}$ to $C^z_{i,j,k}$ can be obtained 
straightforwardly from Eq.~(\ref{eq:Tdiscret}). 
The right hand side is given by
\[
 R_{i,j,k}  =  T^{n}_{i,j,k} + \frac{1}{(c_v \rho)_{i,j,k}} \, Q^{+}_{i,j,k} 
\]

Written in matrix notation Eq.~(\ref{eq:tmat}) reads
\beq
    {\sf M}  \vec{T}^{n+1}  =  \vec{R}
\label{eq:matrix}
\eeq
Obviously the matrix ${\sf M}$ is a sparse matrix with a banded structure.
Usually ${\sf M}$ is diagonally dominant but in situations with extended optically thin
regions this property will be lost.

Equation (\ref{eq:matrix}) can in principle be solved by any linear equation package.
For simplicity and testing purposes we work presently with a standard SOR
solver. Using an optimised relaxation parameter $\tilde \omega$ we need about 130 iterations
per timestep in the initial phase which is far from equilibrium and only 80 iterations
at later times near equilibrium.

The radiation module of the code has been tested extensively in different coordinate
systems in \citet{Bitsch08}, and we have compared our new results
on radiative viscous discs obtained with the 3D version of {\tt NIRVANA} 
in detail with those of the existing 2D code {\tt RH2D}.

\bibliographystyle{aa}
\bibliography{kley8}
\end{document}